\documentclass[prb,twocolumn,groupedaddress]{revtex4}

\usepackage{soul}
\usepackage{physics}

\usepackage{hyperref}

\usepackage{graphicx}

\def\presuper#1#2%
{\mathop{}%
	\mathopen{\vphantom{#2}}^{#1}%
	\kern-\scriptspace%
#2}

\usepackage{environ}
\NewEnviron{NORMAL}{%
    \scalebox{0.85}{$\BODY$}
    }

\usepackage[utf8]{inputenc} 

\usepackage{wrapfig}

\usepackage{tabularx}
\usepackage{capt-of}
\usepackage{color}
\usepackage{subfigure}
\usepackage{multirow}
\usepackage{array}
\graphicspath{{images/}}
\usepackage{amsmath,amssymb,amsfonts,textcomp}
\usepackage{mathtools}

\usepackage{color}
\usepackage{subfigure}
\usepackage{multirow}
\usepackage{array}
\usepackage{outlines}    

\newcounter{ctComment}
\DeclareRobustCommand*{\citen}[1]{%
\begingroup
      \romannumeral-`\x 
      \setcitestyle{numbers}%
      \cite{#1}%
\endgroup
}

\definecolor{dblue}{RGB}{60,105,100}
\definecolor{blue}{RGB}{99,166,159}
\definecolor{yellow}{RGB}{242,225,172}
\definecolor{orange}{RGB}{242,131,107}
\definecolor{pink}{RGB}{242,89,75}
\definecolor{red}{RGB}{205,44,36}

\begin{document}

\title{Cluster many-body expansion: a many-body expansion of the electron correlation energy about a cluster mean-field reference} 

\author{Vibin Abraham}
\author{Nicholas J. Mayhall}
\email{nmayhall@vt.edu}
\affiliation{Department of Chemistry, Virginia Tech,
Blacksburg, VA 24060, USA}

\begin{abstract}
	The many-body expansion (MBE) is an efficient tool which has a long history of use for calculating interaction energies, 
		binding energies, lattice energies, and so on. 
	In the past, applications of MBE to correlation energy have been unfeasible for large systems, 
		but recent improvements to computing resources have sparked renewed interest in capturing 
		the correlation energy using the generalized \textit{n}-th order Bethe-Goldstone equation.  
	In this work, we extend this approach, originally proposed for a Slater determinant, to a tensor product state (TPS) based wavefunction.
	By partitioning the active space into smaller orbital clusters, our approach  starts from a cluster mean field reference TPS configuration and includes the correlation contribution of the excited TPSs using the many-body expansion.
	This method, named \textit{cluster many-body expansion} (cMBE), improves the convergence of MBE at 
		lower orders compared to directly doing a block based MBE from a RHF reference.
	We present numerical results for strongly correlated systems such as the one- and two-dimensional Hubbard models and the chromium dimer.
	The performance of the cMBE method is also tested by partitioning the extended $\pi$ space of several large $\pi$-conjugated systems, 
		including a graphene nano-sheet with a very large active space of 114 electrons in 114 orbitals, 
		which would require $10^{66}$ determinants for the exact FCI solution. 
\end{abstract}

\maketitle

%

\section{Introduction}
Modern electronic structure methods are usually based on the Hartree-Fock reference.\cite{szabo,helgaker2014molecular}
Although most of the energy is already accounted for by this reference, the missing energy, or correlation energy, is necessary in order to obtain accurate and meaningful results.
A full configuration interaction  (FCI)\cite{KNOWLES1984,Knowles1989}
	calculation is required for the exact correlation energy but is unfeasible when
	the system size is large due to its exponential scaling.
Less expensive single reference electronic structure methods like density functional theory (DFT)\cite{Korth2017} or truncated 
	coupled cluster (CC)\cite{shavitt_bartlett_2009} can be used to capture part of this correlation 
	and for most ground state properties.
However, if the system has an ill-defined reference determinant, these methods tend to fail since they are dependent on the reference. 
This type of correlation is broadly referred to as strong or static correlation and usually arises in transition metal complexes, excited states, and bond breaking.
In these cases there is usually orbital near-degeneracy and contributions from more than one determinant becomes important.
One usually resorts to active space based methods for such cases, but even these approaches are plagued by the exponential scaling of the wavefunction with respect to system size.
Improved computational resources and approximations have allowed application of accurate wavefunction-based
	quantum mechanical methods to many challenging strongly correlated systems in recent years.\cite{booth_towards_2013,sharma_low-energy_2014,li_electronic_2019}

The FCI wavefunction is extremely sparse, and there are different approximate methods that can be used to exploit this sparsity.\cite{ivanic_identification_2001}
Selected configuration interaction (SCI) exploits this idea and approximates the wavefunction by selecting important configurations.
The first selected CI algorithm was proposed by Malrieu and coworkers in the 1970's.\cite{Huron1973}
Other selected CI methods include recent improvements to the CIPSI algorithm,
	\cite{Caffarel2016,Yann2017} semi-stochastic heat bath CI,\cite{Holmes2016,Li2018} adaptive CI, \cite{Schriber2016} 
	coordinate descent FCI, \cite{Wang2019} iterative CI, \cite{Liu2016,Zhang2020} adaptive sampling CI \cite{Tubman2016}  and MCCI \cite{Greer1995}.
	The recently proposed full configuration interaction quantum Monte Carlo (FCIQMC) method, samples the determinant basis
	by assigning signed walkers.\cite{Booth2009,Cleland2010}
FCIQMC has also been used to develop the semi-stochastic CAD-FCIQMC where the higher excitations of FCIQMC are used with a CC formalism similarly to externally corrected coupled cluster methods.\cite{Deustua2017,Deustua2018}
A selected coupled cluster method, full coupled cluster reduction (FCCR), has also been proposed and has shown very accurate results with a PT correction.\cite{Xu2018}

Another set of approaches used for solving strongly correlated large active spaces are tensor network based methods.
Density matrix renormalization group (DMRG), \cite{White1992,White1993}
	initially designed for the exact solution of 1D spin lattices, has shown
	impressive results for chemical systems.\cite{White1999,Mitru2001,Chan2002}
DMRG is mostly applicable for pseudo-one-dimensional systems. 
There are also a few higher-dimensional tensor network based methods such as
	tree tensor network states (TTNS), \cite{Nakatani2013} complete graph tensor
	network states (CGTNS)\cite{Marti2010}  and so on. 

Similar to the approach that we will discuss below, 
there are also approaches in which the active space is partitioned into orbital
	groups and then the system is solved by restricting the excitations between those groups.
Occupation restricted multiple active space (ORMAS), \cite{Ivanic2003,Ivanic2003a} restricted active space (RAS)\cite{Olsen1988} and generalized active space (GAS) \cite{Ma2011,Manni2013} can all be conceptualized in this way.


Nesbet in the 1960's proposed to use a many-body expansion (MBE) to capture the correlation energy using the \textit{n}-th order Bethe-Goldstone equation.\cite{Nesbet1,Nesbet2,Nesbet3}
The MBE and its variants\cite{Dahlke2007,Dahlke2008,Schmitt2020,Liu2017,Richard2012,Srimukh2019,Liu2019,Bygrave2012,Zhu2016} 
	are versatile tools used in traditional chemistry applications like predicting binding energies, \cite{richard_2014,Joachim2013}
	crystal lattice energies and structures,\cite{beran_new_2015,Liu2020,Yang640,Wen2012,Nanda2012,Muller2013}
	dipole moment and polarizability,\cite{Medders2013,Peyton2019} 
	vibrational frequencies,\cite{Howard2013,Heindel2018,varandas_adjusted_1995}
	forces\cite{Wang2018,Bates2011,Omar2016,Omar2017}
	and excited state energies.\cite{Liu2019,Jin2020,Paz2021}
Even though MBE has been used in these contexts, its ability to solve for the correlation energy of large systems was not widely exploited
	until recently.
In recent years, with increased computing power and smart pruning, use of the MBE method to approximate FCI energy has seen new interest.
One of the earliest methods where the MBE was used is the method of increment
	(MoI) approach by Stoll where orbital blocks were used as n-body
	entities.\cite{Stoll1992,Stoll2005,Stoll2009,Stoll2010}
Paulus and coworkers have further used the MoI with localized orbitals to study a variety of systems and have also proposed a multi-reference version for bond breaking problems.\cite{Paulus2004,Paulus2006,Fertitta2018}
Ruedenberg and co-workers proposed the correlation energy extrapolated many-body
	expansion where they combined the correlation energy extrapolation by intrinsic
	scaling method with the many-body expansion using local orbitals.\cite{Bytautas2010,boschen_correlation_2017}
The incremental FCI (iFCI) method by Zimmerman \textit{et al.}\cite{zimmerman_incremental_2017,zimmerman_strong_2017} used SHCI \cite{Holmes2016} as a solver for higher order calculations 
	and has also been extended to do orbital optimization.\cite{zimmerman_evaluation_2019,Dang2021}
Eriksen and Gauss proposed the many-body expansion full configuration interaction (MBE-FCI) method by expanding over virtual orbitals\cite{eriksen_virtual_2017,eriksen_many-body_2018,eriksen_many-body_2019}.
A generalized MBE-FCI was later proposed \cite{eriksen_generalized_2019} and has been extended to excited states.\cite{Eriksen2020a}
Recently the incremental approach has also been used with frozen natural orbitals for reducing the dimension of the virtual space dimension at each order.\cite{verma2020scaling}

\begin{figure*}
        \includegraphics[width=\linewidth]{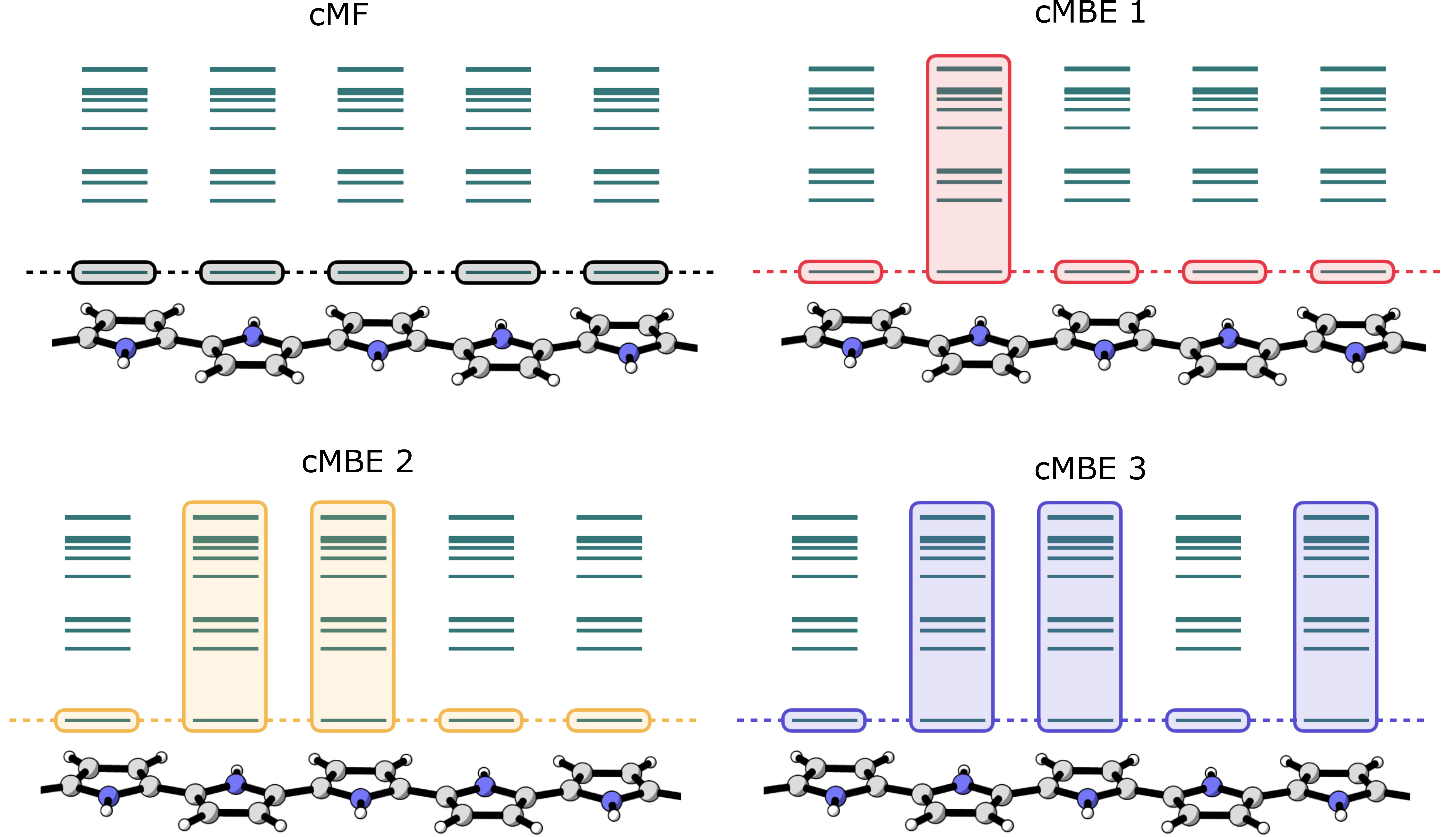}
        \caption{Pictorial depiction of the reference cMF state and example terms from a given subsequent cMBE expansion for the polypyrrole molecule.
		The green lines correspond to the cluster state energies of each cluster (here each pyrrole unit is considered as a cluster).
		The cMF reference as shown is  a single TPS formed by the direct product of the lowest energy cluster states.
		The subsequent many-body expansion can be understood as including the degrees of freedom for the active clusters. 
		We show the example terms for cMBE1, cMBE2 and cMBE3. 
		}

        \label{fig:mbe}
\end{figure*}

\begin{figure}
        \includegraphics[width=\linewidth]{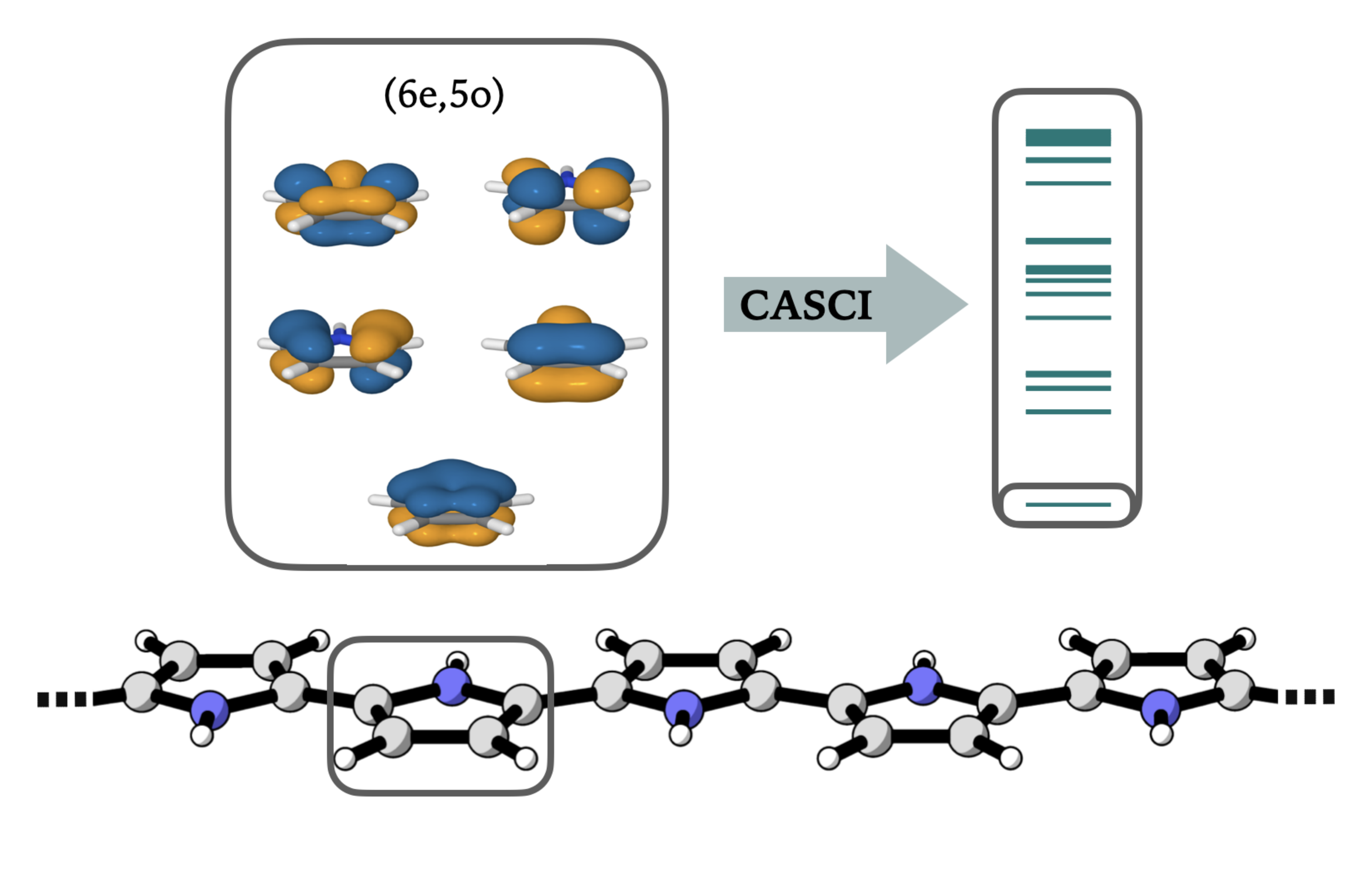}
        \caption{The orbitals within a given cluster are plotted on the left. 
        Solving the exact CASCI problem inside the cluster gives the cluster states in the right panel. The lowest energy cluster state is optimized during cMF.
		}
        \label{fig:basis}
\end{figure}

As the degree of strong correlation increases, higher body corrections need 
	to be incorporated to get exact results since the HF reference
	is not a reliable guess.\cite{eriksen_generalized_2019,zimmerman_strong_2017} 
The perfect pairing reference,\cite{Beran2006} instead of a Hartree-Fock reference has shown improved results for strongly correlated systems\cite{zimmerman_strong_2017}
	but generalizing it to cases where there are more than two orbitals in a block is challenging.
Another drawback of the traditional MBE approaches expanded over an orbital basis is that as the size of the system increases, the calculation of each 
	order gets more expensive with increased number of virtual and occupied orbitals.

In this paper, we propose the use of the tensor product state (TPS) basis as an
	alternative to the traditional Slater determinant basis in the many-body expansion.
The basic idea is to partition the system into separate clusters, solve the smaller many electron problem within each
	cluster, and then represent the wavefunction as a tensor product of these cluster states.\cite{Abraham2020}
Using this alternate wavefunction expansion has multiple advantages when the system is \textit{"clusterable"}.
A system is said to be \textit{"clusterable"} when the Hamiltonian has a structure to it which can be exploited to partition the orbital space into
	different orthonormal orbital clusters.
Even though orbital locality is the most sensible clustering criteria,
	sometimes other factors like symmetry and even bonding and anti-bonding orbital pairings can also lead to
	efficient clustering criteria.\cite{Abraham2020} 
Most of the large molecular systems of interest like crystal lattices or polymetallic complexes have an intrinsic structure which can be used
	to partition the orbitals into separate groups.
Recently Jim{\'{e}}nez-Hoyos and Scuseria proposed the cluster mean field (cMF) method for fermions, where a single
	TPS configuration is variationally minimized.\cite{Hoyos2015}
cMF defines a reference TPS configuration, like Hartree-Fock is the reference determinant for Slater determinant based methods.
There are different methods that use the basic structure of the TPS basis and add
	correlation to it using perturbation theory (PT) \cite{Hoyos2015},
	configuration interaction \cite{Mayhall2017,Abraham2020}, coupled cluster \cite{liBlockcorrelatedCoupledCluster2004,fangBlockCorrelatedCoupled2007,Liu2019b}, 
	tensor networks \cite{parkerCommunicationActivespaceDecomposition2013,parkerCommunicationActiveSpace2014} and effective Hamiltonians.\cite{AlHajj2005a,Ma2012,Morrison2014,Morrison2015}

In the approach described in this paper, the cluster many-body expansion (cMBE), 
	the increments or the building blocks are the clusters themselves and not the virtual or occupied orbitals.
The definitions of the reference state and expansion orders change when using many-electron cluster states as the basis. 
This tensor product state and corresponding orbitals are optimized using cMF.\cite{Hoyos2015} 
Hence in this new representation, the correlation component within each cluster is exactly captured.
Since the exact solution of the cluster basis is used, the size of the clusters is restricted, but with selected CI methods this could also be alleviated.
Using the cMBE method, we propose to both surpass the computational challenge and improve the convergence of the MBE by exploiting the structure of the system.

Some of the advantages of using a mean field TPS reference are: 
\begin{enumerate}
	\item The cMF reference with orbital optimization provides a better reference than Hartree-Fock since it incorporates more correlation.
	\item Faster convergence means fewer terms to compute, which avoids numerical precision issues.
	\item The tensor product reference allows one to exploit localized structure in the molecule for increased efficiency.
	\item Although less of a black-box approach, a cluster based method can provide a more intuitive framework for chemists to analyse and interpret \textit{ab intio} results.
\end{enumerate}

Even with these advantages, the tensor product based methods  do not entirely remove the issue of dimensionality. 
In the tensor product basis, calculating the matrix elements for the Hamiltonian is more expensive than just using the Slater-Condon rules.
One of the attractive features of cMBE is that the higher body corrections can be formed in either the TPS basis or the orbital basis using an effective embedding approach.
We briefly describe the cMF method along with the orbital optimization and the cMBE method in Section \ref{sec:theory}.
We then show the performance of the cMBE method by applying it to the strongly correlated 1D and 2D Hubbard models in Section \ref{sec:hubbard}.
In Section \ref{sec:cr}, we study the commonly benchmarked strongly correlated dichromium system.
In Section \ref{sec:pah}, we apply the cMBE method to the delocalized polycyclic aromatic hydrocarbons (PAH).
We also study the benzene molecule using the cc-pVDZ basis in Section \ref{sec:benzene}.
Finally in Section \ref{sec:conclusion}, we summarize the results and discuss future directions.


\section{Theory}\label{sec:theory}
Traditional wavefunction-based methods start from a mean field determinant and expand the wavefunction as excitations from this reference.
Usually the Hartree-Fock determinant is used as the reference wavefunction and contributions from the excited determinants need to be included for
	chemical accuracy.
The Hartree-Fock orbitals are extremely delocalized and may not provide the best reference orbitals for larger systems.
There are also studies in which CCSD natural orbitals or localized orbitals are used for the MBE.\cite{eriksen_generalized_2019,eriksen_virtual_2017}
Even with these modified orbitals, the reference determinant is not changed significantly and hence can be non ideal for strongly correlated systems.
We propose to use a cMF reference instead of a HF reference and expand the MBE in a TPS basis.
Using this modified reference can be helpful because part of the strong correlation is already included inside the cluster, 
	and the interaction outside can be captured using MBE or any other approach.

To understand the clustering and the TPS wavefunction, we look at the polypyrrole (\textbf{PPy}) polymer unit shown in Figure \ref{fig:mbe}.
This molecule has strong local interactions and weak inter-pyrrole interactions in its ground state and has applications in molecular switches.\cite{Pernaut2000,milczarek_renewable_2012}
In the neutral form, it is a good test system to show the applicability of cMBE.
Each pyrrole unit is a single cluster with 5 orbitals corresponding to the $\pi$ space in each unit.(Figure \ref{fig:basis})
The exact solution of each unit can be solved since it is just a (6e,5o) active space and corresponding cluster states can be generated as shown in Figure \ref{fig:basis}.
In Figure \ref{fig:mbe}, for each unit, the energies of the ground and excited cluster states within each cluster have been plotted.
For the reference, we form a tensor product of each of the ground states in each cluster.
The many-body expansion can then be formed on top of this reference configuration.
In the next section we give a brief description of the cMF method and then introduce the cMBE method in detail.

\subsection{Cluster Mean Field}

The cMF method, originally proposed by Jim\'enez-Hoyoz and Scusceria, is an ideal reference for any TPS based method,
	and has shown promising results for the 1D and 2D Hubbard systems.\cite{Hoyos2015}
In this approach the active space is partitioned into separate clusters or blocks
	and many-electron states are formed within these clusters.
The full system wave function can then be represented using a tensor product of the local cluster states,
\begin{equation}
\ket{\psi} = \sum_{\alpha,\beta,\gamma..\omega} c_{\alpha_{I},\beta_{J},\gamma_{K}..\omega_{N}} \ket{\alpha_{I},\beta_{J},\gamma_{K},..\omega_{N}}  
\end{equation}

where $c_{\alpha_{I},\beta_{J},\gamma_{K}..\omega_{N}} $ corresponds to the coefficient for a given state in the tensor product basis. 
Here we use upper case letters for representing blocks or clusters of orbitals, and
	Greek letters to represent the many electron cluster-states.

The ground state of the full system can be approximated by taking the lowest many-body cluster-state as
\begin{equation}
\ket{\psi_0} =  \ket{0_{I},0_{J},0_{K},..0_{N}}  
\end{equation}
where $\ket{0_L}$ is the lowest energy cluster state for cluster $L$.
We can write the cluster state $\ket{0}_L$ as linear combination of determinants in the cluster $L$.

\begin{equation}
\ket{0_L} = \sum_l x^L_{l,0} \ket{D^l_L}
\end{equation}
where $l$ is the determinant index in the cluster state basis.

The Hamiltonian in the clustered form can be represented as,
\begin{equation}
\hat{H} = \sum_{I}  \hat{H}_I  + \sum_{I<J} \hat{H}_{IJ} +\sum_{I<J<K} \hat{H}_{IJK} + \sum_{I<J<K<L} \hat{H}_{IJKL}
\end{equation}
 where $\hat{H}_I$, $\hat{H}_{IJ}$, $\hat{H}_{IJK}$ and $\hat{H}_{IJKL}$ correspond to Hamiltonian terms with one, two, three and four cluster interactions, respectively.

Analogous to Hartree-Fock,\cite{szabo} we seek a self-consistent optimization of the cluster states $\ket{0_{L}}$ such that the reference TPS is variationally minimized.
The Lagrangian under the constraint that the reference state is normalized can be written as
\begin{equation}
\mathcal{L} = \mel{\psi_0}{\hat{H}}{\psi_0}  + \epsilon \left(\vphantom{\sum} \bra{\psi_0}\ket{\psi_0} - 1\right)
\end{equation}

Differentiating this Lagrangian with respect to the cluster basis coefficients for a given cluster state $\ket{0_L}$,

\begin{equation}
	\label{eq:lag}
	\frac{\partial\mathcal{L}}{\partial{\bra{0_L}}} = 0, 
\end{equation}

and substituting $\ket{\psi_0}$ and the Hamiltonian in clustered form in Equation \ref{eq:lag} yields:
\begin{align}
	\label{eq:diff}
	\frac{\partial}{\partial{\bra{0_L}}}\left(\sum_{M\neq L}\mel{0_L,0_M}{\hat{H}_{L}+\hat{H}_{M}+\hat{H}_{LM}}{0_L, 0_M} \right.\nonumber\\
	\left.\vphantom{\sum_{M}}- \epsilon (\braket{0_L}{0_L}-1)\right) = 0 .
\end{align}



In Equation \ref{eq:diff}, since the differentiation is with respect to cluster $L$, only Hamiltonian terms that have contributions from cluster $L$ are needed. 
Because Hamiltonian terms with three or four body terms ($H_{IJK}, H_{IJKL}$) will necessarily have an odd number of creation/annihilation operators on at least one cluster, they do not contribute to the cMF energy. 
This can be easily demonstrated using an example:
\begin{align}
        \hat{H}_{IJK} \Leftarrow& \sum_{pr\in I}\sum_{q\in J} \sum_{s \in K} \mel{pq}{}{rs}
        \hat{p}^{\dagger}\hat{q}^{\dagger}\hat{s}\hat{r} \\
        &=\sum_{pr\in I}\sum_{q\in J}\sum_{s \in K} \mel{pq}{}{rs}
        \bigg\{\hat{p}^{\dagger}\hat{r} \bigg\} \bigg\{\hat{q}^{\dagger}\bigg\}   \bigg\{\hat{s} \bigg\}
\end{align}
Action of this term on the reference would produce a new TPS with a new electron configuration where cluster $I$ 
    will have same electrons as before, but cluster $J$ will have an extra electron because of the $\hat{q}^{\dagger}$ term and cluster $K$ will have one less electron because of the $\hat{s}$ term.
Hence, three and four body terms do not contribute to the energy evaluation at the cMF step. 


After the differentiation and collecting terms, we have:
\begin{equation}
	\label{eq:fock}
	\left(\hat{H}_L + \sum_M \hat{V}_{L[M]} + E_M \right)  \ket{0_L}  - \epsilon_L \ket{0_L}  = 0, 
\end{equation}
where $\hat{V}_{L[M]} = \mel{0_M}{\hat{H}_{LM}}{0_M}$ is the potential from the cluster $M$ and $E_{M}$ = $\mel{0_M}{\hat{H}_M}{0_M}$.
Equation \ref{eq:fock} is an eigenvalue problem where $\epsilon_{L}$ corresponds to the cluster state energy similar to the orbital energies in HF.
For the fermionic Hamiltonian, $\hat{V}_{L[M]}$ can be represented as 
\begin{equation}
\label{eq:eff}
\hat{V}_{L[M]} =  \sum_{pr\in L}\hat{p}^{\dagger}\hat{r}  \sum_{qs \in M}    \mel{pq}{}{rs} \rho^{M}_{qs}  
\end{equation}
where $\rho^{M}_{qs} = \mel{0_M}{\hat{q}^{\dagger}\hat{s}}{0_M}$ is a one-particle density matrix for cluster $M$. 



Because the effective potential in cluster $L$ has contributions from each cluster $M$ through its one-particle density matrix ($\rho^{M}$), 
we must solve for the cMF state self consistently by updating the effective potential iteratively until convergence.

The Hamiltonian from Equation \ref{eq:fock} can be understood as the many-electron Fock-like operator for the cMF procedure,
\begin{equation}
\hat{H}^0 =\sum_{I} \hat{H}_I  + \sum_{I,J} \hat{V}_{I[J]}.
\end{equation} 

Because these equations arise from a variational minimization of a well-defined energy functional, 
we can easily improve the ansatz by minimizing with respect to the orbitals as well as the cluster state CI coefficients.
As previously demonstrated,\cite{Hoyos2015,Abraham2020} the orbital optimization is a key step that can improve the energy significantly. 

For a given wavefunction $\ket{\Psi_0}$, we can define the unitary transformation in the single particle basis which minimizes the energy 
	using an anti-Hermitian matrix $ \hat{\kappa}$.

\begin{equation}
{\hat{\kappa}}=\sum_{p<q}\kappa_{p q} ({p}^{\dagger} {q}- {q}^{\dagger} {p} )
\end{equation}
where $\kappa_{p q}$ are the orbital rotation parameters.

The single particle basis gets transformed into a new basis.
\begin{equation}
\tilde{\hat{p}}=e^{\hat{\kappa}} \hat{p} e^{-\hat{\kappa}},
\end{equation}
such that the energy now carries an orbital dependence. 
\begin{equation}
E[{\kappa}]=\left\langle\Psi_{0}| e^{-\hat{\kappa}} \hat{H} e^{\hat{\kappa}}| \Psi_{0}\right\rangle
\end{equation}

The orbitals are optimized when the orbital rotation gradient goes to zero. 
Hence the orbital gradient at each step is formed and a conjugate gradient or BFGS algorithm can be used to optimize the orbitals.
The gradient can be expressed in terms of a generalized Fock matrix similar to traditional quantum chemistry methods.\cite{helgaker2014molecular,Bozkaya2013}

\begin{equation}
	G_{pq} = 2 (F_{pq} - F_{qp})
\end{equation}
where $F_{pq}$ is the generalized Fock matrix 
which can be formed using the one- and two-particle density matrices of the cMF reference,
\begin{equation}
	F_{pq} = D_{pr} h_{qr} + \Gamma_{prst} g_{qrst}.
\end{equation}

Therefore, cMF with orbital optimization is identical to a CASSCF with multiple active spaces.
The orbital optimization can be accelerated by forming the orbital Hessian as well,\cite{helgaker2014molecular,Hoyos2015} but we do not take this approach in the current study.

\subsection{Cluster many-body Expansion}
Although cMF provides an exact description of local correlations, as a direct product of single cluster states, it lacks entanglement between clusters.
To reintroduce inter-cluster entanglement, higher energy TPS configurations need to be included to improve the wavefunction.
In this framework, we define a singly excited TPS as when a single
	cluster is allowed to have multiple cluster states rather than just the ground cluster state.
\begin{equation}
\ket{\psi_{\lambda_L}} = \ket{0_{I},0_{J},..\lambda_{L}, ..0_{N}}  
\end{equation}


For a given single excitation, the matrix element between the reference TPS and singly excited TPS can be written as,
\begin{align}
    \mel{0_I,0_J,..\lambda_L,..0_N}{\hat{H}}{0_I,0_J,..0_L,..0_N} =
    \nonumber\\
    \mel{\lambda_L}{\hat{H}}{0_L} 
    =\mel{\lambda_L}{\hat{H}^{0}_L}{0_L} = 0.
\end{align}
This matrix element is zero for a self-consistently optimized TPS reference due to the cMF stationary conditions.
Hence we can define a generalized Brillouin's \cite{szabo} condition for TPS's:
\begin{align}
\mel{\psi_S}{\hat{H}}{\psi_0} &= 0 & \forall S
\end{align}

A doubly excited TPS would be when two clusters are allowed to have full degrees of freedom (Figure \ref{fig:mbe}).

\begin{equation}
\ket{\psi_{\lambda_{L},\mu_{M}}} =   \ket{0_{I},0_{J},..\lambda_{L},..,\mu_{M} ..0_{N}}
\end{equation}

where $\ket{\lambda_{L}}$ and $\ket{\mu_{M}}$ are two excited configurations in the clusters $L$ and $M$  respectively.
In this work, we introduce an incremental approach, \textit{cluster many-body expansion}, on top of the cMF reference to capture the rest of the correlation energy.
Since cMF captures part of the correlation energy missing from Hartree-Fock, we refer to the correlation energy not captured by the cMF as \textit{inter-cluster correlation energy}.
The general many-body expansion method can be written as,
\begin{equation}	
\label{eq:mbe}
E_{c} =  \sum_{I} \epsilon_{I} + \sum_{J<I} \epsilon_{IJ} + \sum_{I<J<K} \epsilon_{IJK} + ...
\end{equation}
where $E_c$ is the inter-cluster correlation energy instead of the traditional correlation energy.
The two body term can be expanded as
\begin{equation}
 \epsilon_{IJ} = E_{IJ} - E_{I}-E_{J}.
\end{equation}
$E_{IJ}$ is the dimer energy where the two clusters I and J have full degrees of freedom as shown in Figure \ref{fig:mbe}. 
This is equivalent to performing a CASCI-like calculation where the active space is composed of orbitals in clusters I and J, 
    embedded in the 1RDM from the rest of the clusters in their ground state.

A three body correction for clusters $I$, $J$ and $K$ can be written as
\begin{equation}	
\epsilon_{IJK} = E_{IJK} - E_{IJ}-E_{JK}-E_{IK}+E_{I}+E_{J}+E_{K}
\end{equation} 
$E_{IJK}$ is the energy of the TPS wavefunction where three clusters I, J and K have full degrees of freedom as shown in Figure \ref{fig:mbe}.
As can be seen, the computational cost of higher body terms would increase drastically.
If all the clusters have $n$ states each, the 3-body term will have a variational space of $n^3$ in the initial Fock space configuration.
For the system in Figure \ref{fig:mbe}, this would be around $10^6$ TPS configurations.
This will become intractable at higher orders very quickly.

One way to tackle this problem is by using a truncated basis in each cluster.
However truncating the cluster basis can affect the final energy quite a lot, especially for systems that have non negligible interactions between clusters.
As a significant improvement over energy-based trunctation, we can instead choose states which are highly entangled via the embedded Schmidt truncation introduced in our previous work.\cite{Abraham2020}
However, even though this can significantly reduce the number of necessary states, the full dimension formed using the tensor product of the states of each cluster will still grow exponentially.
Recently we proposed the tensor product selected configuration interaction (TPSCI) method which approximates the exact solution as a variational linear combination of tensor product states which are chosen by a selected CI procedure. 
TPSCI is ideal for our current purposes since it will adaptively form the wavefunction depending on the interaction between clusters.

One issue with the TPS based approaches is the expensive matrix element evaluation compared to Slater determinant based approaches.
This can be considered as one of the advantages of the cMBE approach compared to other TPS based approaches since the expansion can be computed by avoiding the TPS basis altogether.
For example, for a dimer term $E_{IJ}$, we can compute the effective integrals inside a combined cluster $IJ$ by combining the two clusters.
Hence if cluster I has $n_I$ electrons and cluster J has $n_J$ electrons, 
	the combination of the two clusters forms a new cluster (IJ) with $n_I +n_J$ electrons.
This is similar to forming the effective Hamiltonian inside an active space in a CASCI calculation.
The effective Hamiltonian inside the new cluster can be formed and the CASCI problem can be solved in the determinant basis.
Even for the orbital basis approach where we avoid the TPSs, the new combined active space can be large for higher order terms and approximate approaches ultimately have to be used.
This can be solved using any approximate FCI method, such as selected CI or DMRG or even CCSD(T) if the CAS space becomes large.
Although both methods can be used, it is difficult to tell a priori which one will be ideal for computing higher MBE terms for a general system.
If a system can be clustered efficiently, the TPSCI approach offers unique advantages arising from the natural representation which mirrors the physical system.
Therefore, we use the orbital basis approach for smaller sized cluster systems since CASCI is cheaper and use the TPSCI based approach for large clusters.

Similar to other many-body expansion methods, 
	the approximate dimer and trimer terms can be computed using any many-electron method.
If the system is not fully strongly correlated, traditional methods like CCSD(T)  can  be used to solve for cMBE terms.
For example, if we have a molecular crystal, coupled cluster is a good ansatz for the ground state of each monomer
	as well as the full system.
Hence a CCSD(T) result can be achieved using a many-body expansion with CCSD(T) results for dimer and trimer systems. 
We can even start from a cMF reference with an approximate CCSD density in equation \ref{eq:eff} and use it as the reference for the MBE.
Even though these are interesting possible directions where cMBE can give very convergent results, in this study we only focus on the correlation energy.
The cMBE method is also exactly size extensive even though it is not variational.


As with any adaptive model, the computational complexity of cMBE varies with each application and is rather difficult to precisely characterize. 
Since the method starts with cMF, there is an initial factorial scaling with cluster size, $N$. Assuming the worst-case scenario where each cluster of the $k$ clusters is half-filled, and where exact diagonalization is performed for each of the $n$-body terms, then truncation at $n$ will
have the following scaling:
\begin{align}
    \mathcal{O}\left({k \choose n} {Nn \choose \tfrac{Nn}{2}}^2\right)
\end{align}
However, this the worst cast scenario, as we use approximate diagonalization for higher-body terms (reduces second factor), and use efficient screening to avoid computing negligible $n$-body terms (reduces first factor).

\section{Results}\label{sec:results}
In this section we present data for the cMBE method for a variety of systems.
First we study the half-filled one- and two-dimensional Hubbard model.
We then apply the cMBE method on the strongly correlated dichromium system and some polycyclic aromatic hydrocarbon (PAH) systems.
The PAH systems are extensively delocalized and hence can be considered challenging for the cMBE approach.
Finally, we apply the cMBE method on the recently bench-marked benzene molecule with cc-pVDZ basis.\cite{Eriksen2020}
The integrals for the molecular systems were computed using PySCF package.\cite{pyscf2020}
All DMRG calculations provided for the Hubbard model in this work have been carried out using the \texttt{ITensor} package.\cite{ITensor}

\subsection{Hubbard Model}\label{sec:hubbard}
In this section, we study the one and two dimensional Hubbard model\cite{hubbard1963} using the cMBE method.
The Hubbard Hamiltonian used in this study can be represented using two different hopping values:
 \begin{align}
         \hat{H}=& \sum_{\langle i,j \in A \rangle \sigma}-t_{1} a_{i\sigma}^{\dagger}
 a_{j\sigma}+\sum_{\langle i \in A,j \in B \rangle \sigma}-t_{2}
 a_{i\sigma}^{\dagger} a_{j\sigma} \nonumber \\
         &+U \sum_{j} n_{i \uparrow} n_{j \downarrow}
 \end{align}
 where $t_1$ ($t_2$) are hopping terms within (between) clusters, and $U$ is the same-site Coulomb repulsion.
The Hubbard model becomes strongly correlated when the two-electron Coulomb repulsion ($U$) is much larger than the hopping term.
For all calculations, we use $U = 5$ and $t_1 = 1$ which is strongly correlated regime. 
We study the effect of \textit{clusterabilty} of the system by scanning the $t_2$ hopping term with respect to the $t_1$ parameter.
For example, cMF would be exact for $t_2 = 0$ since the clusters are  non-interacting.

\subsubsection{1D chain}
For the 1D system considered, we present results for values of $t_2:t_1$ = $1:1$, $3:4$ and $1:2$ for a 40 site periodic Hubbard system.
We divide the system into 10 four site clusters.
Since this is a 1D system, we used DMRG for computation of all of the higher order terms in the cMBE method.
We used the original local orbitals and did not perform any orbital optimization since orbital optimization can remove some of the sparseness leading to more terms.\cite{Abraham2020}

\begin{figure}
        \includegraphics[width=\linewidth]{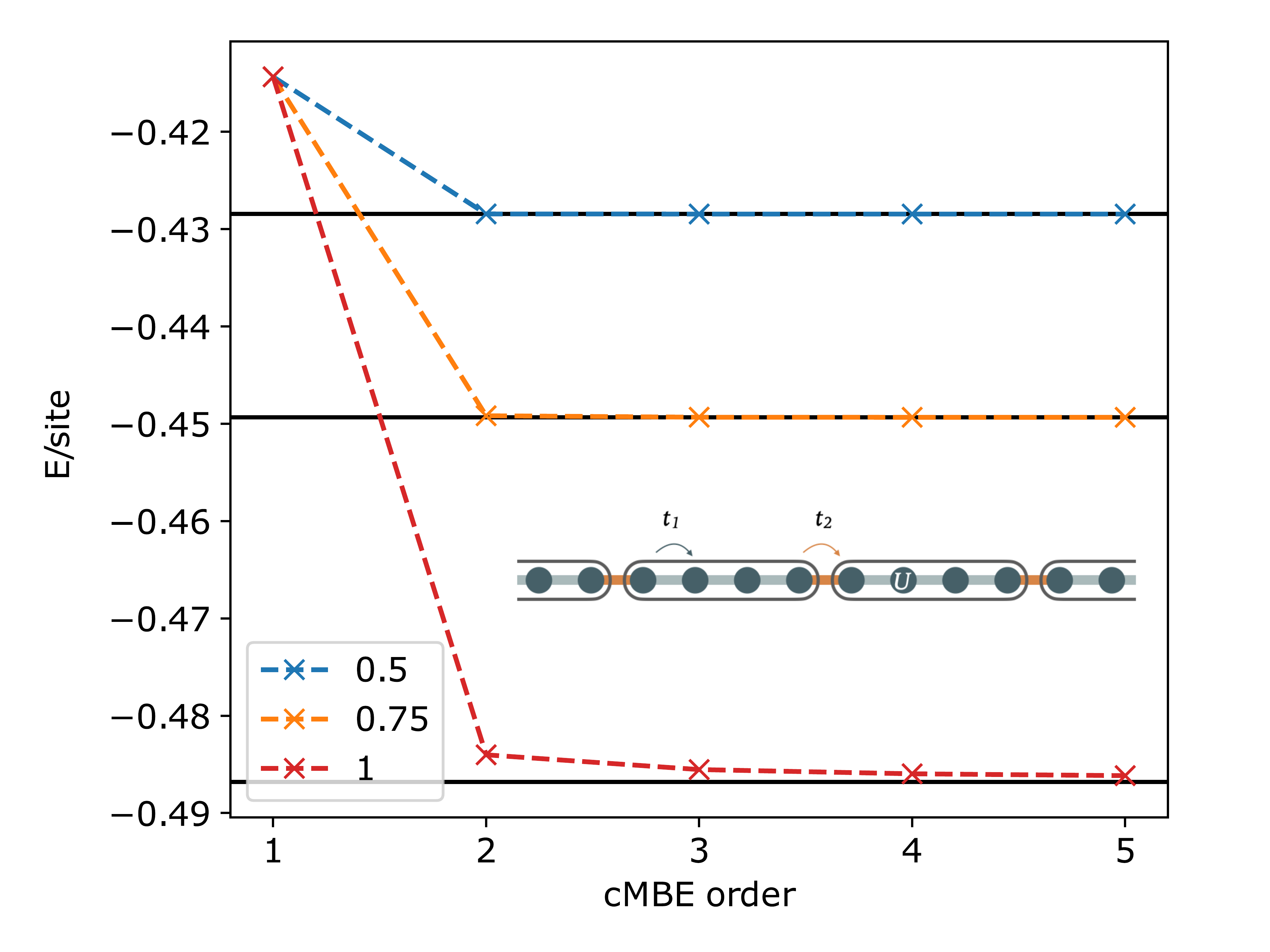}
        \caption{The cMBE energy per site for the 40 site Hubbard model. The black line corresponds to DMRG results.
        Blue line: $t_2/t_1=.5$.  
        Orange line: $t_2/t_1=.75$.  
        Red line: $t_2/t_1=1$.  
                }
        \label{fig:data-hubbard-1d}
\end{figure}

Even though the 1D system is exactly solvable using DMRG, the cMBE gives us a good indication of \textit{clusterability} of these systems. 
From Figure \ref{fig:data-hubbard-1d} it can be seen that for all cases except for $t_2:t_1 = 1:1$, the cMBE expansion converges quickly, almost at second order, even at $t_2:t_1=3:4$.

\subsubsection{2D lattice}
Here we attempt to study the two dimensional Hubbard model with 64 sites as an example. 
We have 16 clusters with 4 sites each.
We consider $t_2:t_1$ =  $1:8$, $1:4$ and $1:2$ for these systems.
For the cMBE results, we  use TPSCI to solve for higher body corrections.
We use the \texttt{FermiCluster} package developed by our group for the cMBE results.\cite{FermiCluster}
The two-dimensional Hubbard model has been studied previously using an
	increment based approach and has shown promising results just by using third order corrections.\cite{hubbard_mbe}
	
\begin{figure}
        \includegraphics[width=\linewidth]{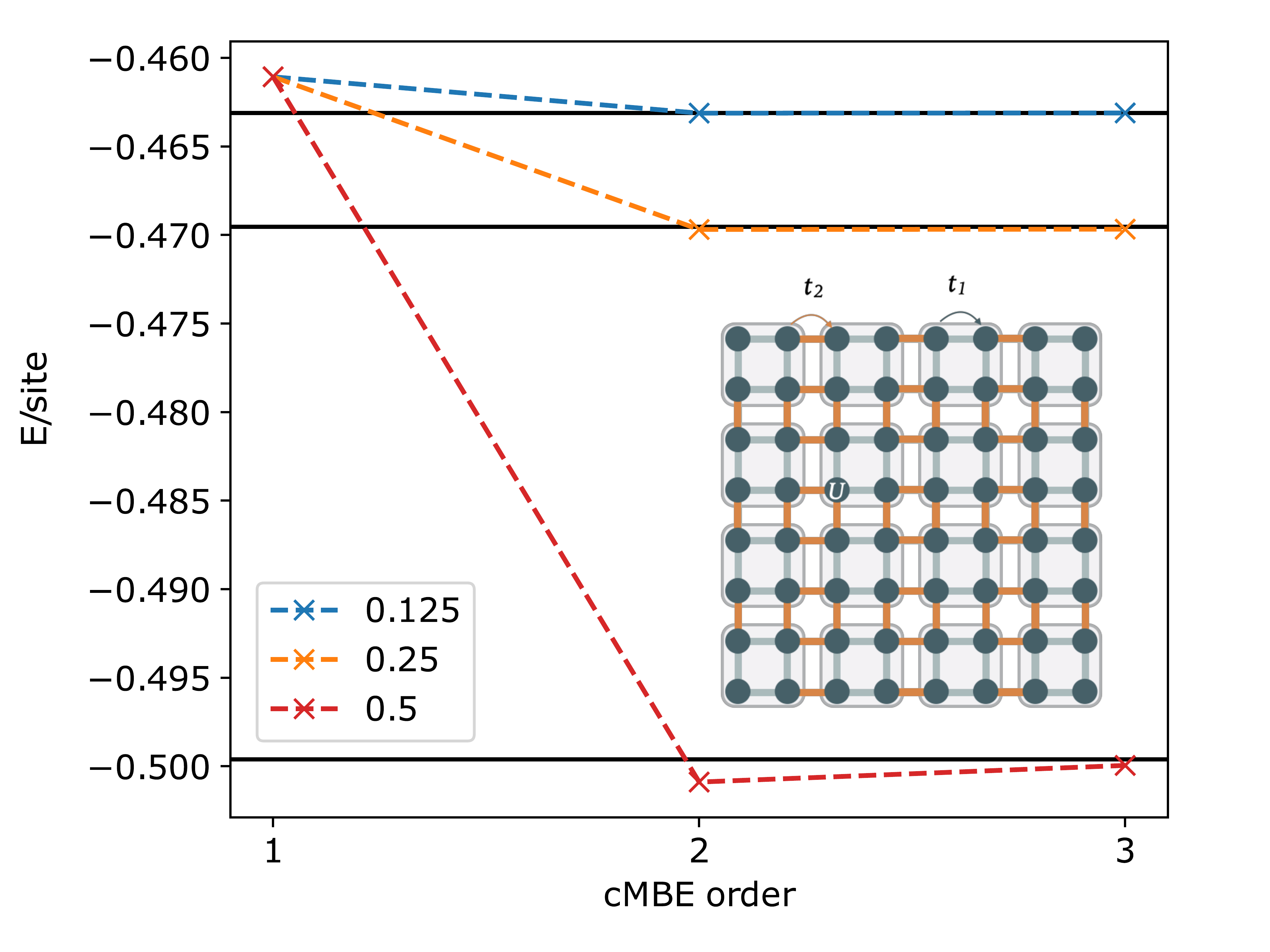}
        \caption{The many-body expansion for the 64 site 2D Hubbard model. The black line corresponds to reference DMRG result with M=1600 for $t_2:t_1$ of $1:8$ and $1:4$ and M=3000 for ratio of $1:2$.
        Blue line: $t_2/t_1=.125$.  
        Orange line: $t_2/t_1=.25$.  
        Red line: $t_2/t_1=.5$.  }
        \label{fig:data-hubbard-2d}
\end{figure}

From the results shown in Figure \ref{fig:data-hubbard-2d}, we can see that for a ratio like $1:8$, the cMBE approach converges quickly.
For these two dimensional systems, the DMRG results also get complicated as we go to higher ratios.
We use the variational TPSCI results for the terms at each order.
The DMRG values are computed using M=1600 except for the case where $t_2:t_1$ = $1:2$ where we use M = 3000.
From Figure \ref{fig:data-hubbard-2d} it can be seen that the cMBE and DMRG values match well for all ratios considered.
We can conclude from the results that the cMBE method can be used for strongly correlated systems with reasonable inter cluster interactions.

\begin{figure*}
        \includegraphics[width=1.0\linewidth]{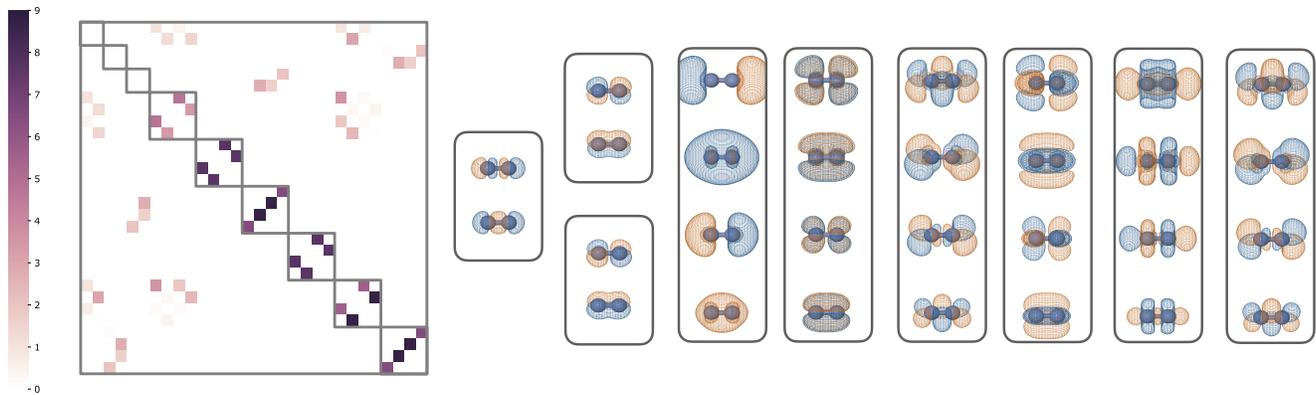}
	\caption{The clustering of the dichromium system. 
		The absolute value of the exchange matrix is plotted and the clusters are selected as blocks. 
		The ordering of the orbital is similar to the ordering of orbitals in the right. 
		Each 4 orbital cluster corresponds to a bond in the hextuple bonded Cr dimer.}
        \label{fig:cr_mo}
\end{figure*}
\subsection{Chromium Dimer}\label{sec:cr}
The chromium dimer at $1.5 \AA$ using an Ahlrichs-SV basis set\cite{sv1992} is a common test system used to study methods developed for strong correlation.\cite{Olivares2015,Yuki2009,Booth2014,Manni2013}
All orbitals up to 3s orbitals are frozen, leading to an active space with 24 electrons in 30 orbitals.
This system is studied to benchmark new methods developed for static and dynamic correlation, since it has a hextuple bond with one $\sigma$, two $\pi$ and three $\delta$ bonds
	involving the 4s orbitals and the 3d valence orbitals.
There are benchmark results computed using DMRG \cite{Olivares2015,Yuki2009}, FCIQMC \cite{Booth2014} and SHCI \cite{Li2020} among others.
In a recent study, Lehtola and coworkers showed that there are excitations as high as octuples using their cluster decomposition method.\cite{Lehtola2017}
The Cr$_2$ has also been studied using the MBE-FCI approach,\cite{eriksen_generalized_2019}
where it was observed that the many-body expansion converges at approximately the 10th excitation rank.


The cMBE approach requires us to partition the orbital space of the system into clusters. 
In the case of the chromium dimer, this may initially seem difficult.
A simple clustering approach is to use each bond as a separate cluster. 
As shown by the exchange matrix in Figure \ref{fig:cr_mo}, the $3d$ orbitals and $4d$ orbitals with
	the similar shape have large off-diagonal elements, implying significant interaction. 
Hence, we put these $3d$ and $4d$ orbitals in the same cluster.
The  bonding and antibonding orbital pairs formed from each atom's $3p_x$, $3p_y$, and $3p_z$ atomic orbitals, which are fully occupied, are each treated as their own cluster. 
The orbital clustering (shown in Figure \ref{fig:cr_mo}) is: 

(3p$_z$), (3p$_x$), (3p$_y$), (3s, 4s), (3d$_{xy}$, 4d$_{xy}$), (3d$_{yz}$, 4d$_{yz}$), (3d$_{xz}$, 4d$_{xz}$),(3d$_{x^2-y^2}$, 4d$_{x^2-y^2}$), (3d$_{z^2}$, 4d$_{z^2}$)


We present the data for the Cr dimer in Table \ref{tab:cr2}.
Coupled cluster with up to quadruple exictations is unable to get a good estimate for this system.\cite{Olivares2015}
The cMBE result for the 4 body correction is within chemical accuracy compared to other FCI quality results.\cite{Booth2014,Yuki2009}
We use the tightly converged TPSCI+PT results for the individual terms for the cMBE.
We use the selection threshold 1e-10 and search threshold 1e-3 for the variational space.

cMBE or any  other TPS based methods are ideally designed for studying spatially extended molecules where localized orbitals can be used
	and the drastic decay of electron correlation can be taken advantage of.
The presented Cr dimer data demonstrates that the cMBE approach, even though not meant or designed for small systems such as diatomics, is capable of providing very accurate results.
Given the notably non-monotonic convergence of the energies with MBE order, a definitive assessment will require more accurate calculations incorporating 5-body terms into the calculations. This is needed to confirm that the cMBE calculation has actually converged on the accurate point (especially since the 4-body terms provided a significant effect).
However, we were not able to converge the individual 5-body terms accurately enough with the current implementation. 
In future work we plan to further exploit the cluster basis representation to simplify such higher-body terms. 

\begin{table}[]
\caption{Correction at each order up to four-body correction for the Cr$_2$ system (24e,30o). The orbital basis used is RHF. We present reference values for other methods using the same HF core.}
\begin{tabularx}{0.7\linewidth} {
				  >{\centering\arraybackslash  \hsize=.5\hsize}X
			|	  >{\centering\arraybackslash  \hsize=.5\hsize}X
				   }
\hline
\hline
\multicolumn{1}{c|}{Order} & \multicolumn{1}{c}{cMBE} \\ \hline 
1                           & -2085.9921               \\ 
2                           & -2086.4482               \\ 
3                           & -2086.3278               \\ 
4                           & -2086.4211               \\ \hline 
\hline
  Method                          &  Energy              \\ 
\hline
CCSDTQ\cite{Olivares2015}   & -2086.4067               \\ 
DMRG\cite{Yuki2009}         & -2086.4211               \\ 
FCIQMC\cite{Booth2014}	    & -2086.4212		\\ 
 SHCI\cite{Li2020}	    & -2086.4211		\\ \hline
\hline
	\end{tabularx}
\label{tab:cr2}
\end{table}

\begin{figure*}
        \includegraphics[width=1.0\linewidth]{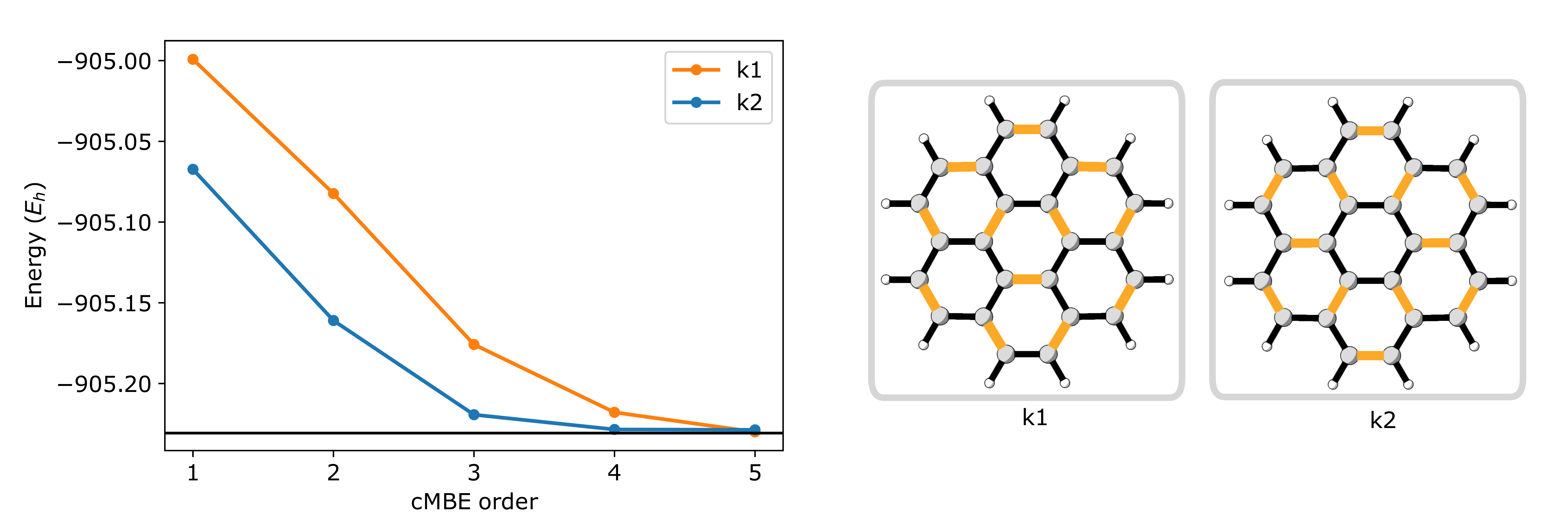}
        \caption{Comparison of the two different clustering options considered for the coronene molecule. The highlighted yellow region corresponds to the atom pairs in a cluster.}
        \label{fig:coronene}
\end{figure*}

\subsection{PAH system}\label{sec:pah}
Next, we study several polycyclic aromatic hydrocarbon (PAH) systems using the cMBE method.
Due to the extended $\pi$ conjugation in these systems, they have possible applications in light-emitting diodes, solar cells and so on.
Large graphene nanodots \cite{Chen2018} have applications in bioimaging and photovoltaics
	and can be considered as strongly correlated systems.
The extended $\pi$-conjugation in these systems makes them a relatively hard problem for a fragmentation based approach like cMBE.
Hence the PAH systems should be challenging for the cMBE method.
Geometries for all the PAH systems in this study were optimized using B3LYP with the cc-pVDZ basis except for the hexacene system whose geometry was obtained from Ref. \citen{Hachmann2007}. All structures are provided in the Supporting Information. 

There exist multiple classical rules which give a good qualitative idea of the correlation and structure of PAH systems.\cite{masic2006,Alonso2010,yeh_role_2016}
One such model is the Kekulé structure.
The PAH molecules are mainly formed by conjugated double bonds, hence it makes sense to cluster PAH systems using 
	one of its Kekulé structures as the guiding clustering option.
Because this leads to a two orbital cluster, we can include higher order corrections to see the convergence properties of the cMBE approach.

\subsubsection{Different Kekulé structures}
For a simple PAH system, there can be multiple Kekulé clusterings possible.
In this section, we study the coronene molecule using two different clustering approaches.
It has been previously debated whether the coronene molecule has weak outer
	double bonds or two concentric $\pi$-conjugations, leading to clustering in
	\textbf{k1} or \textbf{k2} respectively in Figure \ref{fig:coronene}.
	By defining a clustering based on a Kekul\'e structure (whereby the $p_z$ orbitals defining each double bond in a given Kekul\'e structure are taken to be a cluster),
cMBE can be used to determine which clustering provides a more physically correct picture.
Based on just the Kekulé structure, it is difficult to say which of the two clustering is ideal.
We present data for both clustering approaches in Figure \ref{fig:coronene}.
Both Kekulé clusters provide the same Clar's structure in the end, but it can
	be seen that \textbf{k2} clustering is more convergent than \textbf{k1}.
We also note that we found the cMF reference to have a lower energy for
	\textbf{k2} compared to \textbf{k1} which can be used as a good indicator for
	relatively better clustering.
Using this as a metric, we can avoid doing the expensive cMBE for all possible Kekulé structures.
This is usually the case with systems which do not have too many empty virtual orbital clusters.
It has been previously demonstrated experimentally and theoretically that the
	coronene molecule has weak outer double bonds unlike benzene which clearly
	suggests that the \textbf{k2} clustering would be better.\cite{Nikita2018,Sulflower2017}
Even with \textbf{k2}, we need to go to higher orders for the expansion to converge since the system is delocalized.

\begin{figure*}
        \includegraphics[width=\linewidth]{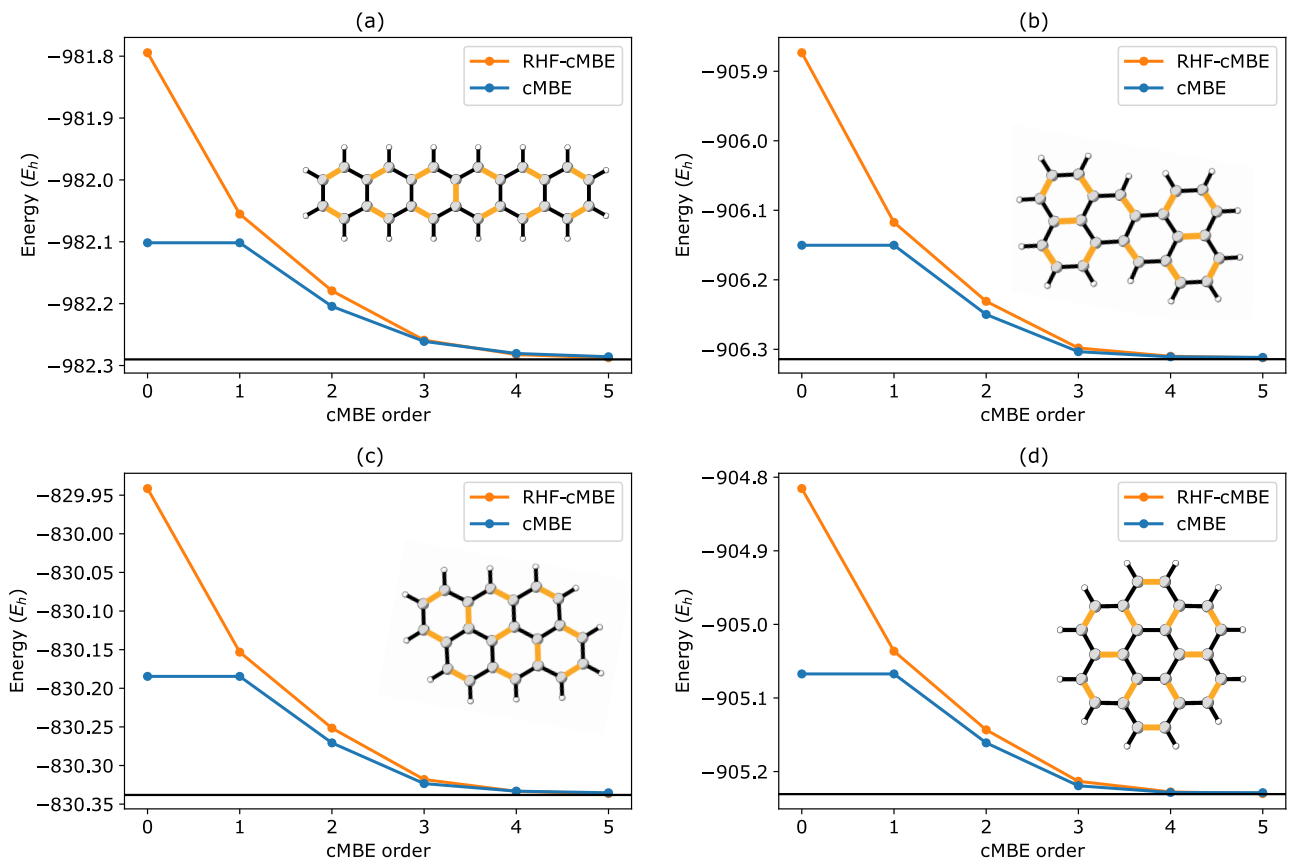}
        \caption{Comparison of cMF reference with using split localized RHF basis for four PAH systems. 
        The yellow highlighted region of the molecules corresponds to the double bonds considered as clusters. 
        The reference value corresponds to the DMRG result for (a) and extrapolated SHCI results for (b),(c) and (d).}
        \label{fig:cmf_rhf}
\end{figure*}

\subsubsection{Comparison between cMF and split localized RHF basis}
Now, we move on to evaluate the effect of starting from a cMF reference by comparing it with the RHF reference. 
As mentioned, the MBE can be formulated without the cMF reference by directly using RHF orbitals.
We take a few PAH molecules and localize the occupied and virtual $\pi$-orbitals separately.
We then cluster the orbitals based on bonding/antibonding pairs.
This leads to an automatic stable Kekulé type clustering for some PAH systems.
In contrast to the cMF reference, where part of the correlation is already captured, the MBE using RHF reference starts from a less stable RHF solution. 
For the RHF based MBE, the reference is the RHF determinant with all the occupied orbitals doubly occupied. 
The first order correction for this RHF-cMBE amounts to a CAS calculation for a single cluster while constraining the occupied orbital in all other clusters to be doubly occupied.
The two-body correction is then a CAS calculation by combining the orbitals in the two clusters while others are doubly occupied and so on.

We present data for the comparison of the cMBE with the split localized RHF-cMBE method in Figure \ref{fig:cmf_rhf}.
It can be seen that the cMF reference is much better than the RHF reference. 
The cMBE and the RHF-cMBE approach converge to the exact result at higher orders
	with cMBE having better convergence for all the systems.
The reference energies are extrapolated SHCI values computed using the \texttt{Arrow}\cite{Arrow} package except for the hexacene molecule, for which we use the DMRG value from Ref. \citen{Hachmann2007}.
Pruning can be used for truncating the number of terms for larger systems.
In the Supporting Information, we present two different pruning techniques, \texttt{scheme-0} and \texttt{scheme-1} and apply it to the Kekulene molecule. 


\begin{figure*}
        \includegraphics[width=\linewidth]{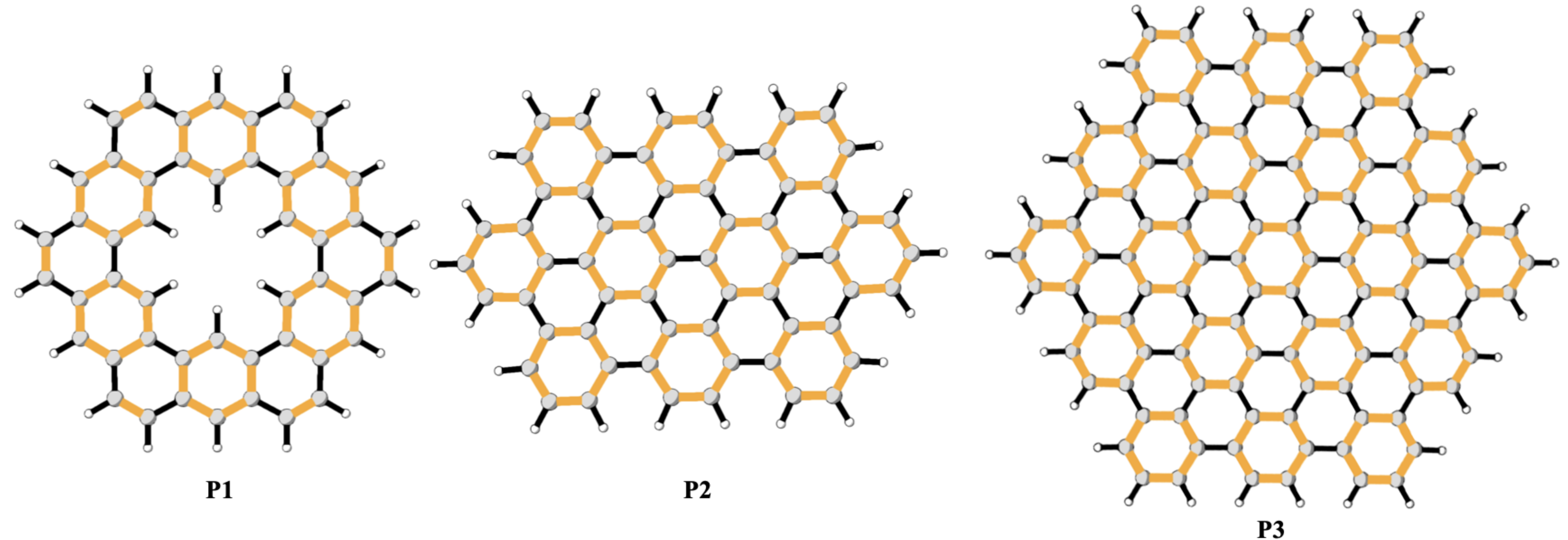}
        \caption{Large PAH systems considered in this work using the Clar's rule clustering. The active space comprises of the $\pi$-conjugated electrons. The yellow highlighted regions corresponds to a single cluster.}
        \label{fig:pah_systems}
\end{figure*}

\begin{table*}[]
\caption{Data for the large PAH systems studied using the Clar's rule clustering.}
\begin{tabularx}{\textwidth} {
				  >{\centering\arraybackslash  \hsize=.15\hsize}X
				| >{\centering\arraybackslash  \hsize=.17\hsize}X
				| >{\centering\arraybackslash  \hsize=.17\hsize}X
				| >{\centering\arraybackslash  \hsize=.17\hsize}X
				| >{\centering\arraybackslash  \hsize=.17\hsize}X
				| >{\centering\arraybackslash  \hsize=.17\hsize}X
				   }
\hline
\hline
   & \multicolumn{1}{c|}{CCSD(T)} & \multicolumn{1}{c|}{Extrp SHCI} & \multicolumn{1}{c|}{cMF} & \multicolumn{1}{c|}{cMBE2} & \multicolumn{1}{c }{cMBE3} \\ \hline
   \textbf{P1} (48e,48o)   & -1810.3857                & -1810.3842                    & -1810.2330              & -1810.3578              & -1810.3838  \\
   \textbf{P2} (60e,60o)   & -2258.5624                & -2258.5360                    & -2258.3885              & -2258.5262              & -2258.5559  \\
   \textbf{P3} (114e,114o) & -4283.9319                &     -                         & -4284.2751              & -4284.5790              & -4284.5999   \\ \hline \hline
	\end{tabularx}
\label{tab:pi}
\end{table*}

\subsubsection{Clar's clustering}
For the PAH systems, even though the Kekulé clustering gives very accurate results, we need to go to larger orders of cMBE.
In this section, we consider a larger cluster size based on Clar's rule such that most correlation is captured within the clusters and the cMBE can be truncated at lower orders.
The Clar's aromatic sextet rule has also been used with a fragmentation based DFT approach and provide promising results.\cite{Noffke2020}
For larger systems, the Clar's rule based clustering would be more ideal.

First we consider the Kekulene molecule.
Kekulene, similar to coronene also has two possible structures: the two superaromatic inner and outer ring or the Clar's sextet based structure.
The Kekulene molecule was recently synthesized and visualized using ultra-high-resolution atomic force microscopy (AFM) and the superaromatic behaviour was not observed.\cite{Pozo2019}
There are computational studies also suggesting Kekulene to have the Clar's
	rule based structure,\cite{Haijun1996} hence we expect more convergent behaviour using the Clar's clustering.


The Kekulene molecule (\textbf{P1}) in Clar's type clustering leads to a 12 cluster system
	with six two orbital clusters and six sextet clusters.
The \textbf{P2} molecule, which is a part of a graphene nanosheet has an active space of 60 orbitals in 60 electrons.
This system requires about $10^{34}$ determinants for the exact results.
We also study an even larger nonographene system \textbf{P3} which has an active space of 114 orbitals in 114 electrons.
The FCI space for this molecule in the $\pi$ space would have $10^{66}$ determinants for the ground state.
This system has 19 clusters and we provide data up to third order correction.

For all of these systems, we use the energy obtained using the TPSCI+PT for the tuples. 
The TPSCI method, being a selected CI approach,  forms a smaller variational space for clusters that interact less.
For example, the interaction between the $\pi$ sextets at the two corners of the \textbf{P3} molecule are nearly negligible.
Hence considering the two body term between these two clusters have a variational space of 94 TPS configurations.
The variational space for one of the nearest neighbour interactions had approximately 4000 configurations.
Both these values are much less than the full dimension for a 12 orbitals in 12 electrons active space which has 853,776 determinants.
Using a RHF reference and performing CAS calculations for the many-body expansion would be intractable at 3rd order.
Hence using a TPSCI procedure, we avoid forming the full space dimension for the two and three body terms for the cMBE approach.
Using the Clar's clustering, the cMBE approach should converge faster and higher order corrections would not be required.
Obtaining higher body corrections would require further coarse-graining of the variational space.
There are multiple possible ways this can be achieved which will be discussed in future work.

In Table \ref{tab:pi}, we present data for the large PAH systems using the Clar's rule based clustering and compare them to extrapolated SHCI and CCSD(T) results.
For the \textbf{P1} molecule, the errors are within $0.2 mE_h$.
For the next largest PAH system, \textbf{P2}, the storage of the PT space gets very large, hence
	the SHCI values could not be computed at very accurate threshold.
Hence the extrapolated SHCI number for this system, as seen from Table \ref{tab:pi} is not a good estimate.
The \textbf{P3} molecule is a larger graphene type system and has extended electron delocalization.
It is interesting to observe that the variational cMF energy is already lower in energy that CCSD(T), even before adding any many-body interactions.

%

\begin{figure}
        \includegraphics[width=1.0\linewidth]{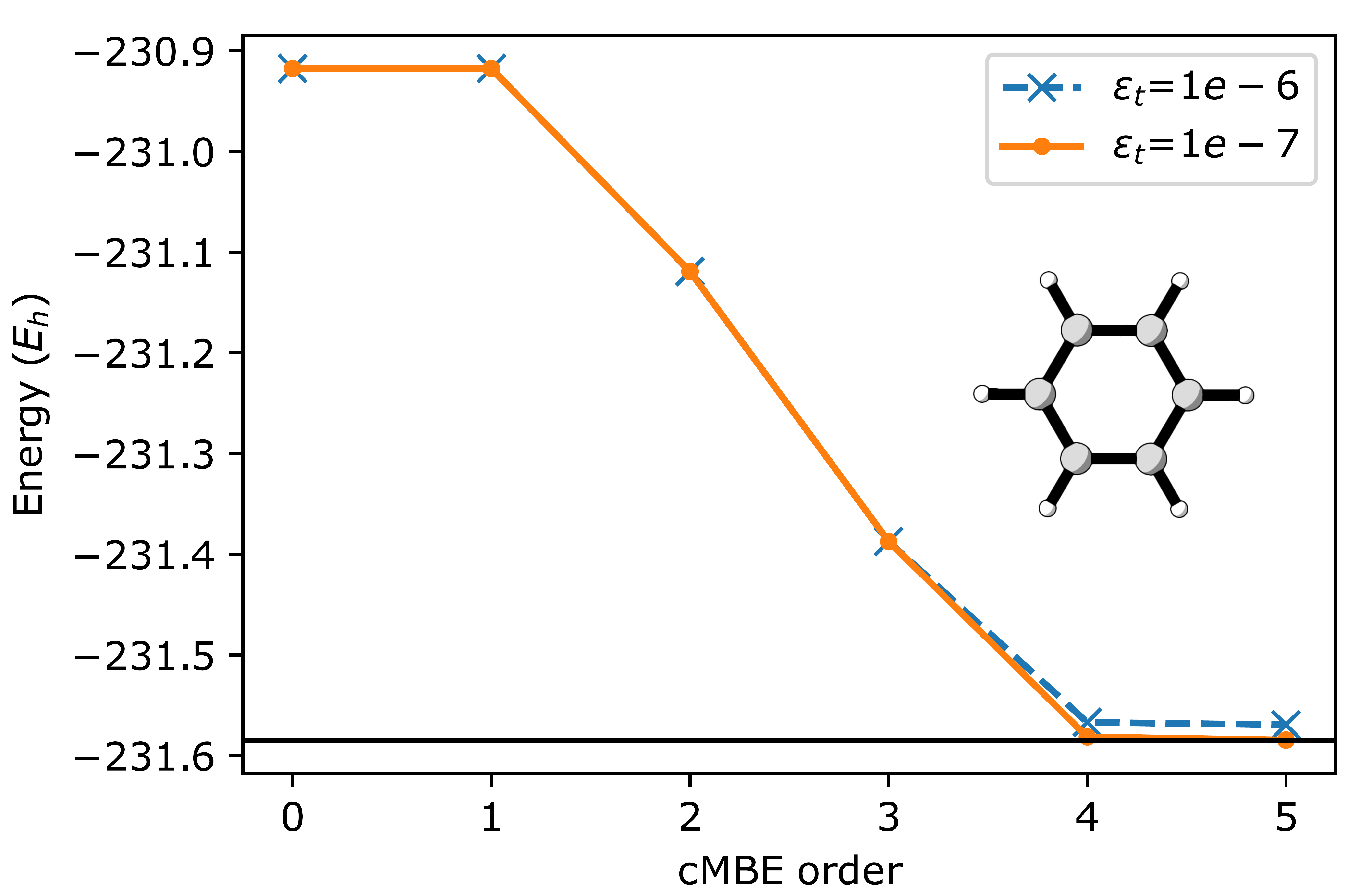}
	\caption{The cMBE values for the Benzene system with cc-pVDZ basis
		using \texttt{scheme-1} pruning. 
	The threhsold used for two different cMBE calculation is also given in the legend.
	The black line corresponds to the estimated energy value from Ref. ~\citen{Eriksen2020}.}
        \label{fig:benz}
\end{figure}

\subsection{Benzene cc-pVDZ}\label{sec:benzene}

In a recent review, most of the methods mentioned in the introduction section
	were tested in a blind challenge to approach the full configuration interaction
	energy of the benzene molecule in a cc-pVDZ basis.\cite{Eriksen2020}
Even though the benzene molecule is not strongly correlated, the numerically exact result for this large active space with 30 electrons in 108 orbitals is not trivial.
The correlation energy for this system was estimated to be  $-863 mE_h$.\cite{Eriksen2020}

There is no simple strategy to cluster the benzene molecule and partitioning the benzene system is a non-trivial problem.
However since the focus of this work is not the optimal clustering but rather the cMBE method itself, we do not yet explore different clustering possibilities.
We localize the $\pi$ space by taking both occupied and virtual orbitals from the $\pi$ bond formed by the $2p_z$ orbitals.
The rest of the occupied and virtual orbitals are localized separately.
We use a clustering where all the $\sigma$ bonds and the corresponding antibonding orbitals are in a single cluster.
The $\pi$ space is then clustered using a Kekulé structure.
Hence, the valence and occupied space are relatively easier to cluster compared to the large virtual space.
For simplicity, we also clustered the full valence space using a Kekulé criteria by pairing the orbitals with similar shape in adjacent C atoms into a cluster.
This basically means that the system is partitioned into three main parts, orbitals corresponding to C1-C2, C3-C4 and C5-C6. 
Within these groups the orbitals are clustered such that orbitals with same principle quantum numbers, angular momentum and shape are in the same cluster.
For example, each of the $3p_x$ orbitals in C1 and C2 are in a cluster and similarly for other orbitals. 
This makes a total of 54 clusters with 6 C-H clusters, 6 C-C clusters and 14$\times$ 3 clusters for the valence $\pi$ space and virtual orbtials.
The clustering for the benzene system is  pictorially depicted in Supporting Information.

For the benzene molecule, we present the data in Figure \ref{fig:benz}.
We use the \texttt{scheme-1} pruning for the benzene system.
All the tuples are considered up to third order. 
We present data for two different pruning thresholds $\epsilon_t$,  \texttt{1e-6} $E_h$ and \texttt{1e-7} $E_h$ .
The $\epsilon_t$ is the pruning cutoff as discussed in the Supproting Information.
For the loose threshold, the correlation energy is not fully captured even at 5 body.
This is expected since we have a large virtual orbital space and we need to include the effect of even smaller contributions.
For the tighter threshold, we get correlation energy of $-862.6$ $mE_h$ at 5 body which is within the range of estimated FCI result.\cite{Eriksen2020,Rask2021}
Only a very small fraction of the terms are considered using pruning.
For the tight threshold calculation, only 16 $\%$ and 2 $\%$ of the total possible
	terms are considered after pruning at 4 body and 5 body, respectively.
The most expensive calculation for the cMBE is the 5 body terms, which still is only a 10 orbital active space and is solvable without much computational resources.

\section{Conclusion}\label{sec:conclusion}
We present a new increment based method on top of a cMF wavefunction formed by clustering the strongly interacting orbitals into clusters. 
We show how the cMBE method can be used to obtain the correlation energy for different types of systems including model Hamiltonians and molecular systems. 
The cMBE approach gives good results for systems with large active spaces as presented in this work.
If only expectation values are needed, it can be used as an alternative to other methods like DMRG and selected CI when the wavefunction is approximately clusterable.
We test this method on large $\pi$-conjugated systems and even the strongly correlated dichromium system.

Future work will focus on further decreasing the computational cost. 
One simple approach could be to use an approximate treatment (e.g., CCSD(T)) for the cluster calculations.  
Another future direction could be to include a PT correction from other clusters on top of the cMBE similar to using RPA or MP correction terms for higher orders in traditional methods.\cite{bygrave_embedded_2012}
This would lead to higher body corrections at every correction level, which can improve the convergence of the expansion. 
We can also introduce a better pruning criteria than the energy based criteria used in this work.
For example, a distance based criteria can also be used for the PAH systems.\cite{Liu2017a,Ouyang2016}

One of the main future directions would be to extend the cMBE method to do larger systems and apply it to other properties like excited states and interaction energies.
Recently the many-body expansion with diabatic states method was proposed and applied to charge transfer reactions.\cite{Paz2021}
A similar framework can be used with a multi-reference cMF to study excited states using the cMBE approach.

Another interesting direction we are interested in is to use the TPS framework and coarse-grain the degrees of freedom at each order. 
This would lead to a compact wavefunction and allow application of the TPS framework for higher body corrections for larger clusters.

\section{Data Availability}
The data that supports the findings of this study are available within the article [and its supplementary material].

\section{Supplementary Material}
Supplementary material is provided containing information about the orbital clustering of benzene, a comparison of a couple pruning techniques, 
and all the xyz structures.

\section{Acknowledgements}
This research was supported by the National Science Foundation (Award No. 1752612). 

\bibliographystyle{achemso}
\providecommand{\latin}[1]{#1}
\makeatletter
\providecommand{\doi}
  {\begingroup\let\do\@makeother\dospecials
  \catcode`\{=1 \catcode`\}=2 \doi@aux}
\providecommand{\doi@aux}[1]{\endgroup\texttt{#1}}
\makeatother
\providecommand*\mcitethebibliography{\thebibliography}
\csname @ifundefined\endcsname{endmcitethebibliography}
  {\let\endmcitethebibliography\endthebibliography}{}

\end{document}


\title{Supporting Information: Cluster many-body expansion: a many-body expansion of the electron correlation energy about a cluster mean-field reference} 

\author{Vibin Abraham}
\author{Nicholas J. Mayhall}
\email{nmayhall@vt.edu}
\affiliation{Department of Chemistry, Virginia Tech,
Blacksburg, VA 24060, USA}

\maketitle

\beginsupplement
\section{Pruning for large active spaces}
We study a larger system, Kekulene, which has a large active space with 48 electrons in 48 orbitals using the conjugated double bond type clustering.
With such a large system, the number of calculations at higher orders becomes very large.
It is prudent to use pruning to cut down the number of calculations.
Pruning of terms is common in MBE based approaches.\cite{eriksen_virtual_2017,eriksen_many-body_2018}
We define two types of pruning techniques to cut down the factorial cost of the number of terms in the cMBE approach.
For example, let us consider a four body term with clusters \texttt{[a,b,c,d]}.
The decision to compute a tuple \texttt{[a,b,c,d]} is made by looking at the tuples \texttt{[a,b,c] [a,b,d] [b,c,d]} and \texttt{[a,c,d]}.
We define the \texttt{scheme-0} pruning, where the tuple \texttt{[a,b,c,d]} is computed if the energy contribution of all the child tuples are above a given threshold $\epsilon_t$.
This is similar to the pruning introduced by Eriksen \textit{et al.}.\cite{eriksen_virtual_2017}
This leads to the use of very tight $\epsilon_t$ values for good accuracy.
We can further relax the criteria and define a \texttt{scheme-1} pruning, where the tuple is computed if N - 1  tuples are important out of the N possible tuples.

\begin{figure}
        \includegraphics[width=0.6\linewidth]{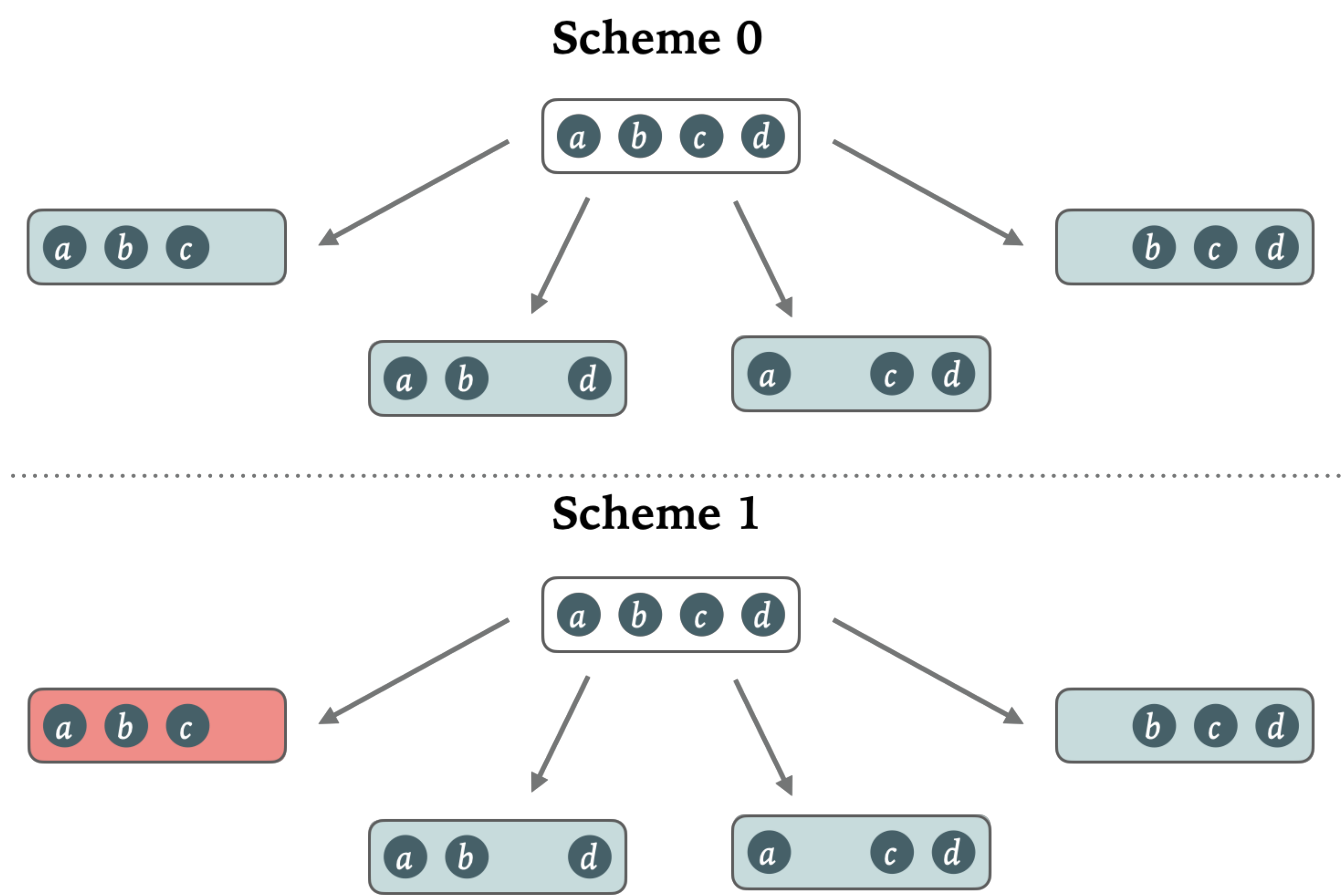}
	\caption{The pruning \texttt{scheme-0} and \texttt{scheme-1} used in
		this work. For \texttt{scheme-0} all the child tuples energy contributions need to be above
		the threshold ($\epsilon_t$) for the parent tuple calculation to be carried out.
	For \texttt{scheme-1}, all except one child tuple's energy contribution needs to be above the threshold.}
        \label{fig:prune}
\end{figure}
\begin{figure*}
        \includegraphics[width=\linewidth]{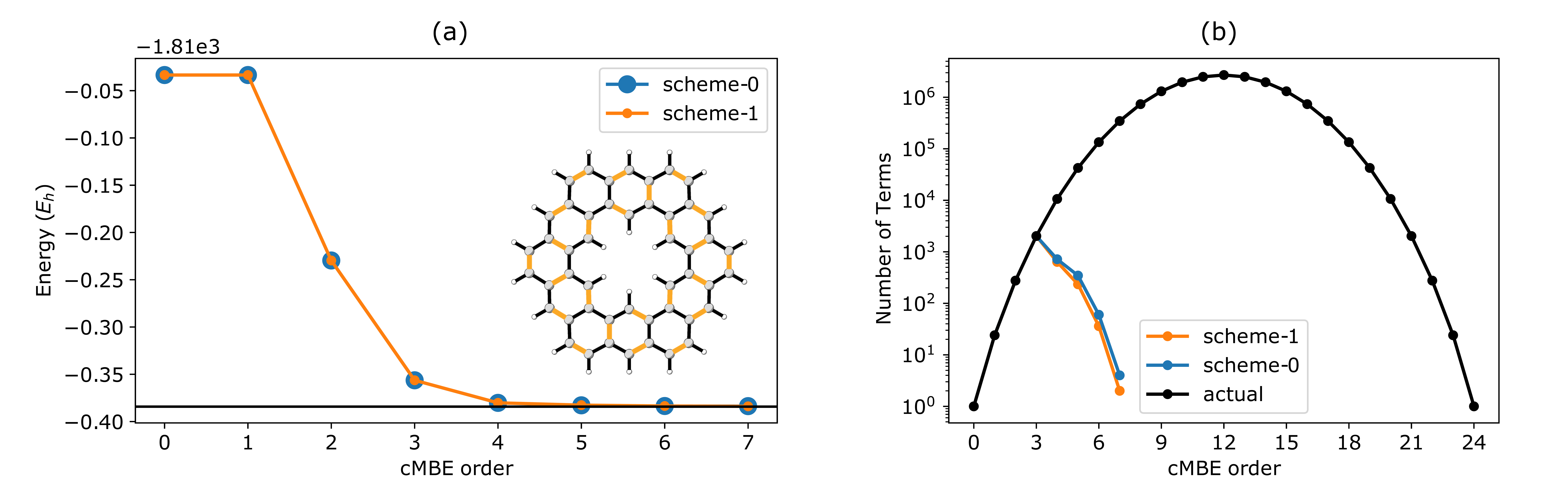}
        \caption{Comparison of the two pruning algorithms for the kekulene molecule using the Kekulé clustering.
	a) the energy per order is plotted with respect to the cMBE order. 
	b) Number of terms computed for \texttt{scheme-1} and \texttt{scheme-0} compared with the \texttt{actual} number of terms needed at each order.
	}
        \label{fig:c48_kekule}
\end{figure*}
We can keep going further to even more relaxed criteria, but the pruning might not be very efficient.
The reason for using a relaxed criteria can be understood by looking at a 1D lattice with three adjacent clusters \texttt{a,b,c}.
With the \texttt{scheme-0} threshold, the tuple \texttt{[a,c]} which is a next nearest neighbour tuple may be below a given $\epsilon_t$ and will get ignored in \texttt{scheme-0}.
However, \texttt{[a,b,c]} is a linear three body term which is important for the final energy.
Using a \texttt{scheme-1} criteria allows one of the three tuples to be negative and hence we can have a sensible truncation.
Both the pruning schemes are pictorially depicted in Figure \ref{fig:prune}.

We study the Kekulene molecule using both the pruning criteria. 
All terms up to three body were considered.
For the \texttt{scheme-0}, we used a pruning threshold $\epsilon_t$ of \texttt{1e-6} $mE_h$ for the 4-body terms. 
For the \texttt{scheme-1} method, a more relaxed pruning threshold $\epsilon_t$ of \texttt{5e-6} $mE_h$ was used.
The threshold was decreased by a factor of 2 for the subsequent terms.
It would be interesting to study the effect of different pruning algorithms on the final energy of the system using the cMBE approach but that is beyond the scope of current study.

Results for the Kekulene system are presented in Figure \ref{fig:c48_kekule}.
The \texttt{scheme-1} method with a relaxed threshold is a bit better than the \texttt{scheme-0} but the errors are within chemical accuracy for both pruning algorithms
	and hence non-distinguishable at convergence.
It can be seen from Figure \ref{fig:c48_kekule} that the number of calculations is much smaller compared to the full expansion and can be computed easily.

\section{Polypyrrole}
\begin{figure}
        \includegraphics[width=0.6\linewidth]{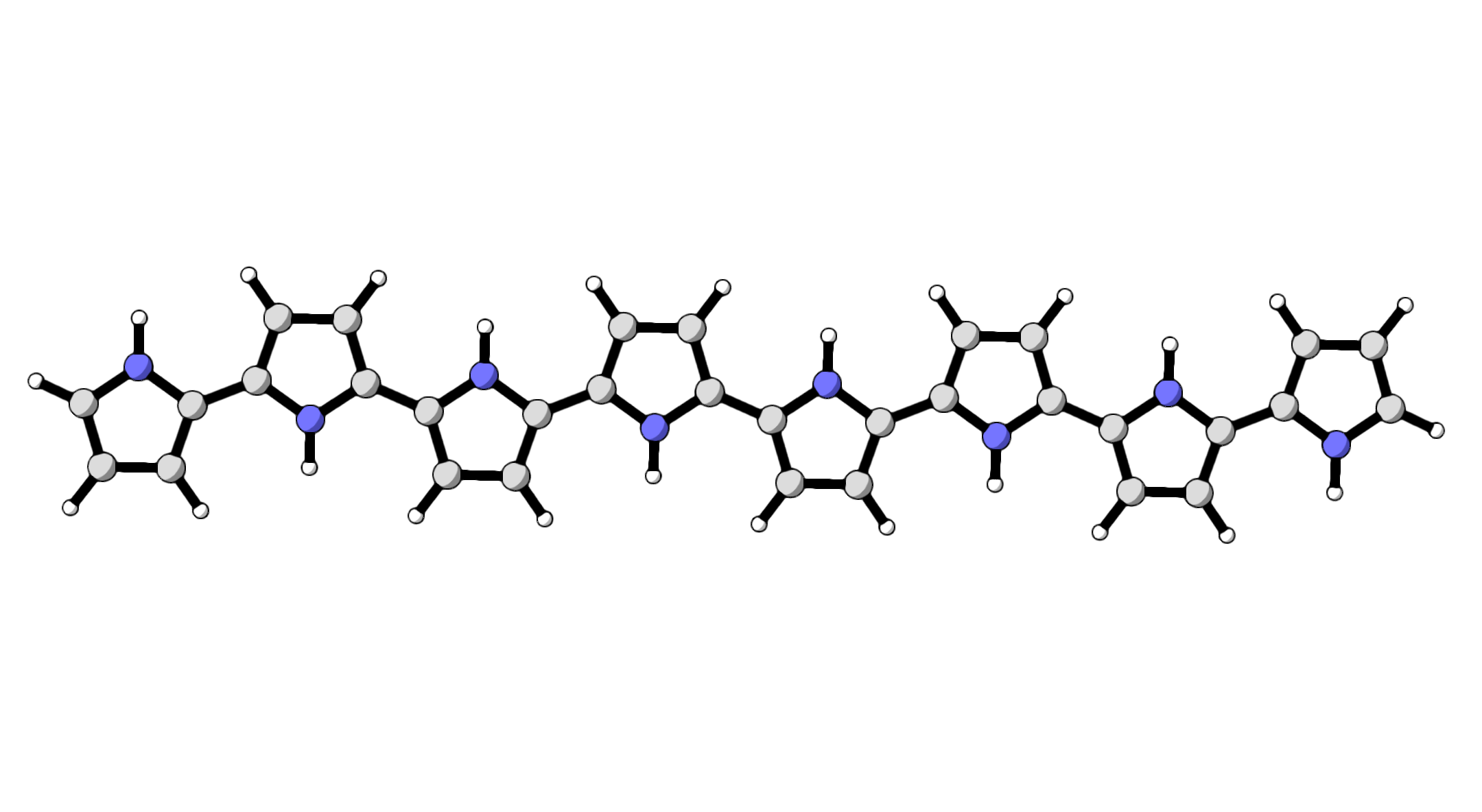}
        \caption{The \textbf{PPy} molecule considered in this study.}
        \label{fig:ppy}
\end{figure}

We also study a polymer unit with a heteroatom.
\textbf{PPy} can be used in biomaterial applications like artificial muscles because of their biocompatible material. \cite{Smela2003,park_heparinpolypyrrole_2007}
Due to its high conductivity and simple synthetic procedure it also has applications to biosensors,\cite{Pernaut2000}  renewable cathodes, \cite{milczarek_renewable_2012} efficient photocatalysts,\cite{yuan_photocatalytic_2019} and supercapacitors.\cite{huang_nanostructured_2016}
Here we present the \textbf{PPy} molecule in its reduced neutral form with 8 pyrrole units as shown in Figure \ref{fig:ppy} as an example.
The active space for a single pyrrole unit is (5o,6e) using $\pi$ space orbitals.
Therefore the full active space for the system becomes (40o,48e).

\begin{table}[]
\caption{Correction at each order up to 3 body correction for the \textbf{PPy} polymer with 8 pyrrole units.}
\begin{tabularx}{0.4\linewidth} {
				  >{\centering\arraybackslash  \hsize=.4\hsize}X
				| >{\centering\arraybackslash  \hsize=.6\hsize}X
				   }
\hline
\hline
\multicolumn{1}{c|}{Order} & \multicolumn{1}{c}{cMBE} \\ \hline 
1                           & -1642.4376                \\ 
2                           & -1642.4878                \\ 
3                           & -1642.4908                \\ \hline
\hline
  Method                    &  Energy                       \\ \hline
CCSD(T)                     & -1642.4896                    \\ \hline
\hline
	\end{tabularx}
\label{tab:ppy}
\end{table}

We present the cMBE results for \textbf{PPy} molecule in Table \ref{tab:ppy}.
We provide CCSD(T) energy for comparison.
The error between the CCSD(T) result and the cMBE value is within 1mH.
In the oxidized bipolaron form \cite{Chen1997} the molecule becomes more delocalized and studying that with a single cMF reference would not be sufficient.
A multi-reference version of cMF can be used in that case.
The cMBE can be defined on top of this MR cMF similar to the work developed by Fertitta \textit{et al.} for Slater determinants\cite{Fertitta2018} and will be investigated in future studies.

\section{Clustering of Benzene molecule}
\begin{figure}
        \includegraphics[width=1.0\linewidth]{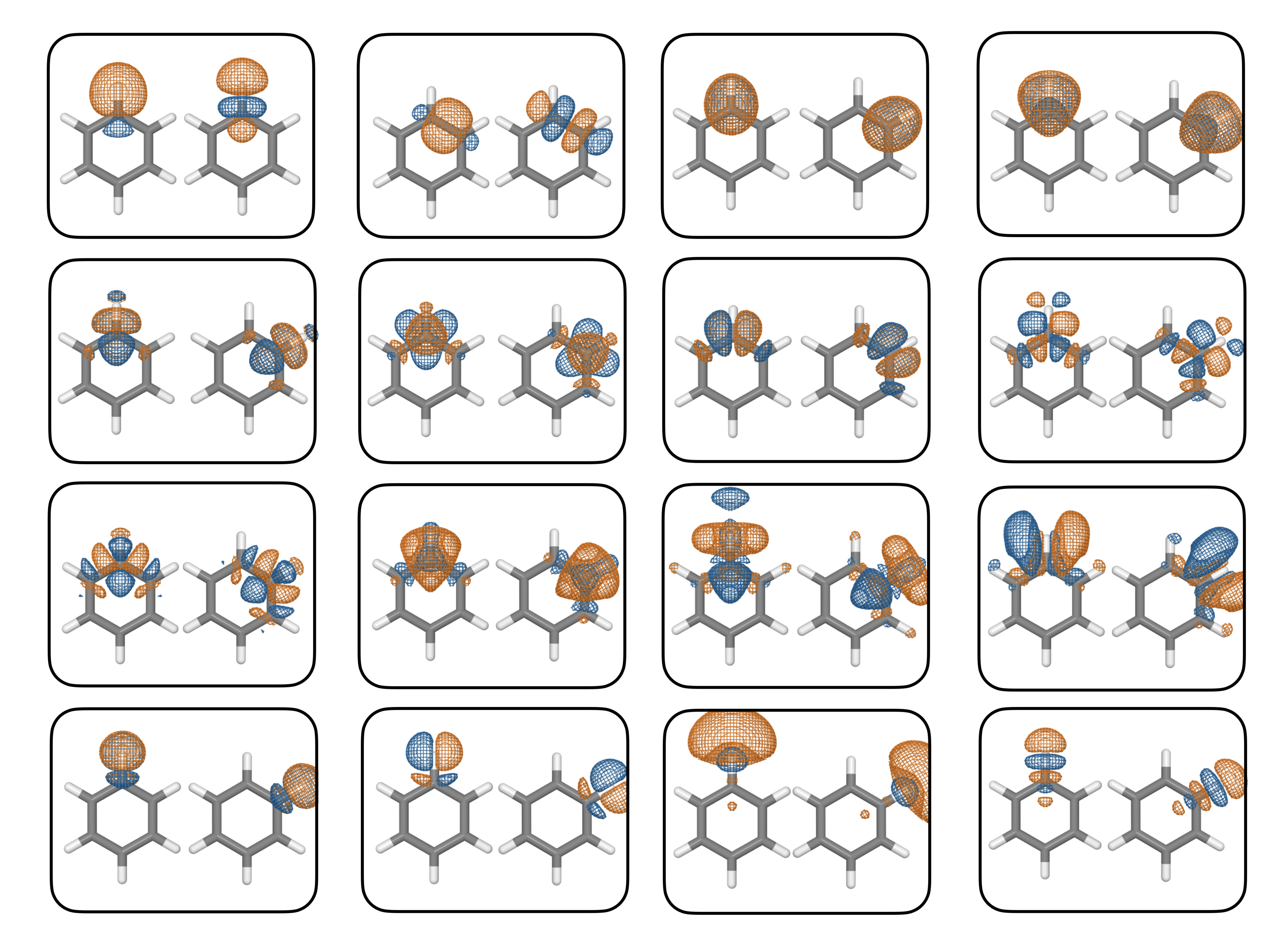}

	\caption{The localized orbitals of benzene using the cc-pVDZ basis for C1-C2
	The C-C and C-H sigma bonding /antibonding  orbitals were taken as a cluster as shown in the first two panels.
	The rest of the valence $\pi$ orbitals and the virtual orbitals were partitioned using the Kekulé structure, where similar orbitals in C1 and C2 are paired. 
	        \label{fig:benz_mo}
	}
\end{figure}

Clustering used for the benzene benchmark is presented in Figure \ref{fig:benz_mo}.
